\documentclass[11pt]{article}
\pdfoutput=1

\usepackage[margin=1in, paperwidth=8.5in, paperheight=11in]{geometry}

\usepackage{graphicx}
\usepackage{float}
\usepackage{subcaption}
\usepackage{fp}
\usepackage{pgfplots}
\pgfplotsset{compat=1.13}

\usepackage{tikz}

\usepackage{verbatim}

\usepackage[numbers]{natbib}
\usepackage{url}

\usepackage{afterpage}

\begin{document}

\title{Stochastic Characteristics of Qubits and Qubit chains on the D-Wave 2X}
\author{John E. Dorband\\
	Department of Computer Science and Electrical Engineering\\
	University of Maryland, Baltimore County\\
	Maryland, USA\\
	\texttt{dorband@umbc.edu}}
\date{\today}
\maketitle

\begin{abstract}
This document presents a studies of the stochastic behavior of D-Wave qubits, qubit cells, and qubit chains. 
The purpose of this paper is to address the algorithmic behavior of execution rather than the physical behavior, though they are related. 
The measurements from an actual D-Wave adiabatic quantum computer are compare with calculated measurements from a theoretical adiabatic quantum computer running with an effective temperature of zero.
In this way the paper attempts to shed light on how the D-Wave's behavior effects how well it minimizes its objective function and why the D-Wave performs as it does.
\end{abstract}

\section{Introduction}\label{sec:intro}

The D-Wave\citep{Dwave13} is an adiabatic quantum computer\citep{Farhi00,Giuseppe08}. The problem class that it addresses is based on the objective function:
\begin{equation}\label{eq:obfunc}
min\left({\sum\limits_i a_i q_i + \sum\limits_i \sum\limits_j b_{ij} q_i q_j}\right)
\end{equation}
where $q_i$ are the qubit values returned by the D-Wave, $a_i$ and $b_{ij}$ are the coefficients given to the D-Wave associated with the qubits and the qubit couplers respectively. 
The D-Wave returns the set of qubit values which minimize the above objective function. Theoretically there is only one global minimum value, even if there are multiple global minima.  
This minimum value corresponds to the ground state of the D-Wave with the given coefficients.  
The D-Wave often returns a non-minimum energy state due to inherent noise in the system, and the closeness of a large number of slightly higher energy active states near the ground state.  
This leads to the question: if the D-Wave does not always return qubit values corresponding to the ground state what are the properties of the D-wave that can be depended upon to perform useful computations. 
The purpose of this paper is to present the stochastic characteristics of the D-Wave that may be useful in understanding quantum adiabatic algorithmic computations.

The D-Wave architecture is base on a non-complete graph for coupling, the chimera graph (see figure \ref{fig:chimera}). 
Thus it is necessary to organize physical qubits into virtual qubits that have greater connectivity than the chimera graph provides for physical qubits. 
This concept is used throughout algorithmic development for the D-Wave. 
Virtual qubits arranged as physical qubit pairs are presented in a paper about solving the Map Coloring problem with the D-Wave\citep{Dahl15}. 
Also virtual qubits organized as physical qubit chains were used to implement on the D-Wave a large complete bipartite graph with higher degree of connectivity than is currently available in the D-Wave\citep{Adachi15}. 
We will present the results of our study of the stochastic properties of virtual qubits, as qubit chains and qubit cells in this paper. 

\section{Measuring Stochastic Properties}\label{sec:metrics}

In this section the stochastic property of a qubit entity whether it is a qubit, a qubit cell or a qubit chain is measured as the probabalility that an individual qubit of the entity 1) is on ($P(q)$), 2) has an up spin ($P(\uparrow)$), or 3) has a value of one ($P(q=1)$).

\subsection{Properties of a Qubit}\label{sec:qubitmetric}

To measure the stochastic property of the D-Wave qubit, the D-Wave was run 10000 times with all the coupler coefficients, $b_{ij}$, set to zero ($C_{c}=0.0$) and all the qubit coefficients, $a_{i}$, set to a specific value, $C_q$, in the range [-1,1]. 
This was done for 129 values over the [-1,1] range. 
Figure \ref{fig:qubitprob} shows the qubit properties of System6 (BS6) and  System13 (BS13) at Burnaby, BC, and C12 (AC12) at NASA Ames Research Center, CA.  

\begin{figure}
    \begin{subfigure}{0.49\textwidth}
	\scalebox{0.75}{
	    \begin{tikzpicture}
		\input{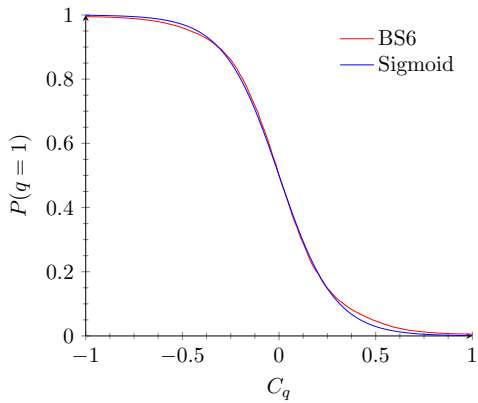}
		  \begin{axis} [
      axis lines=left,
      xlabel={$C_q$},
      ylabel={$P(q=1)$},
      ymin=0, ymax=1,
      xmin=-1, xmax=1,
      minor x tick num=3,
      minor y tick num=3,
      legend cell align=left,
      legend style={ draw=none, },
    ]
    \addplot[red]    table[ x index=1, y index=2 ]   {\plotdata};
    \addplot[blue]   table[ x index=1, y index=7 ]   {\plotdata};
    \legend{BS6,Sigmoid}
  \end{axis}
	    \end{tikzpicture}
	}
	\caption{ } \label{fig:BS6}
    \end{subfigure}
    \hspace*{\fill} % separation between the subfigures
    \begin{subfigure}{0.49\textwidth}
	\scalebox{0.75}{
	    \begin{tikzpicture}
		\input{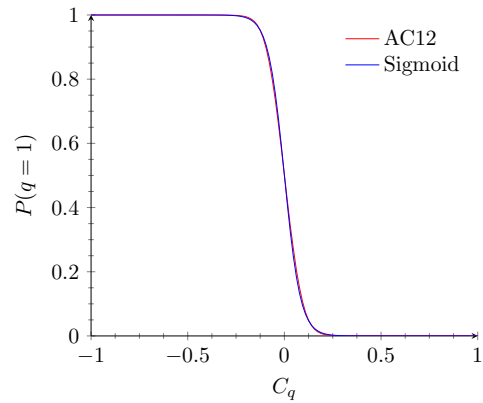}
		  \begin{axis} [
      axis lines=left,
      xlabel={$C_q$},
      ylabel={$P(q=1)$},
      ymin=0, ymax=1,
      xmin=-1, xmax=1,
      minor x tick num=3,
      minor y tick num=3,
      legend cell align=left,
      legend style={ draw=none, },
    ]
    \addplot[red]    table[ x index=1, y index=4 ]   {\plotdata};
    \addplot[blue]   table[ x index=1, y index=8 ]   {\plotdata};
    \legend{AC12,Sigmoid}
  \end{axis}
	    \end{tikzpicture}
	}
	\caption{ } \label{fig:AC12}
    \end{subfigure}
    \caption{The plots of probability of a qubit having a value of 1 overlayed with the plot of the corresponding sigmoid for (a) BS6 [k=7] and (b) AC12 [k=24].}
    \label{fig:sigBS6AC12}
\end{figure}

The probabilities displayed in figure \ref{fig:qubitprob} can be modeled as a function of effective temperature\citep{Adachi15,Benedetti15}:
\begin{equation}\label{eq:effectivetemp}
P(q=1)=\frac{e^{-f(C_q,T_e)}}{e^{f(C_q,T_e)}+e^{-f(C_q,T_e)}}
\end{equation}
where $T_e$ is the effective temperature and $f(C_q,T_e)=\frac{C_q}{T_e}$.
The theoretical behavior of a perfect quantum qubit that would minimize the objective function (Eq. \ref{eq:obfunc}) can be represented by $T_e=0$ or a step function at $C_q=0.0$ as in figure \ref{fig:qubitprob}. 
None of the D-Wave qubit characteristics are as good as the theoretical qubit, since no quantum system can be perfectly isolated from it's environment. Note however that the properties of the BS13 and the AC12 are much better than the older SB6. 
This is probably due to improvements in the newer D-Waves in isolating it from its quantum environment (eg. lower running temperature). 
The equation \ref{eq:effectivetemp} can be reduced to: 
\begin{equation}\label{eq:sigmoid}
P(q=1)=\frac{1}{1+e^{kC_q}}
\end{equation}
where $k={2}/{T_e}$.
Figure \ref{fig:sigBS6AC12} is the comparison of the theoretical model with the measure data where $k=7$ is used for SB6 and $k=24$ is used for AC12.
\begin{figure}
\centering
\begin{minipage}{0.5\textwidth}
	\centering
	\scalebox{1.5}{% A Chimera graph
% Author: John E. Dorband

% Chimera Graph Preamble (required
%\usepackage{tikz}

\begin{tikzpicture}[
  qubit/.style={circle, fill=black, thick, minimum size=2pt},
  scale=0.30,
  inner sep=0pt,
  ]

  \newcount\qu
  \newcount\x
  \newcount\y
  \newcount\xe
  \newcount\xf
  \newcount\ye
  \newcount\yf

  % Add east/west edges between cells
  \foreach \shd in {0,1,...,15} {
      \foreach \shr in {1,3,5} {
	  \draw [red] (\shr,\shd) .. controls (\shr + 0.5,\shd + 0.5) and (\shr + 1.5,\shd + 0.5) .. (\shr + 2,\shd);
      }
  }

  % Add north/south edges between cells
  \foreach \shd in {0,1,...,11} {
      \foreach \shr in {0,2,4,6} {
	  \draw [blue] (\shr,\shd) .. controls (\shr - .75,\shd + 1.0) and (\shr - .75,\shd + 2.5) .. (\shr,\shd + 4);
      }
  }

  \foreach \shd in {0,4,...,12} {
      \foreach \shr in {0,2,...,6} {
	  \foreach \xa in {0,...,1} {
	      \foreach \ya in {0,...,3} {

		  \x=\shr
		  \advance\x by \xa

		  \y=\shd
		  \advance\y by \ya

		  \qu=\x
		  \multiply\qu by 4
		  \advance\qu by \y

		  \node[qubit] (N-\the\qu) at (\x,\y) {};
	      }
	  }
	  \xe=\shr
	  \xf=\xe
	  \advance\xf by 1
	  \foreach \yg in {0,...,3} {
	      \foreach \yh in {0,...,3} {
		  \ye=\shd
		  \advance\ye by \yg
		  \yf=\shd
		  \advance\yf by \yh
		  \draw (\xe,\ye) -- (\xf,\yf);
	      }
	  }
      }
  }

\end{tikzpicture}}
	\caption[ISING DW CvsQ]{Chimera Graph.}
	\label{fig:chimera}
\end{minipage}%
\begin{minipage}{0.5\textwidth}
	\centering
	\scalebox{0.75}{
	    \begin{tikzpicture}
		\input{tab/QubitCharacter}
		  \begin{axis} [
      axis lines=left,
%      title=Qubit Charteristics,
      xlabel={$C_q$},
      ylabel={$P(q=1)$},
      ymin=0, ymax=1,
      xmin=-1, xmax=1,
      minor x tick num=3,
      minor y tick num=3,
      legend cell align=left,
      legend style={ draw=none, },
    ]
    \addplot[red]    table[ x=bias, y=BS6 ]         {\plotdata};
    \addplot[green]  table[ x=bias, y=BS13 ]        {\plotdata};
    \addplot[blue]   table[ x=bias, y=AC12 ]        {\plotdata};
    \addplot[orange] table[ x=bias, y=Theoretical ] {\plotdata};
    \legend{BS6,BS13,AC12,Theoretical}
  \end{axis}
	    \end{tikzpicture}
	}
	\caption{Qubit Characteristics. The probability that a qubit will return a value of 1 given a specific qubit coefficient value $C_q$.}
	\label{fig:qubitprob}
\end{minipage}
\end{figure}

\begin{figure}
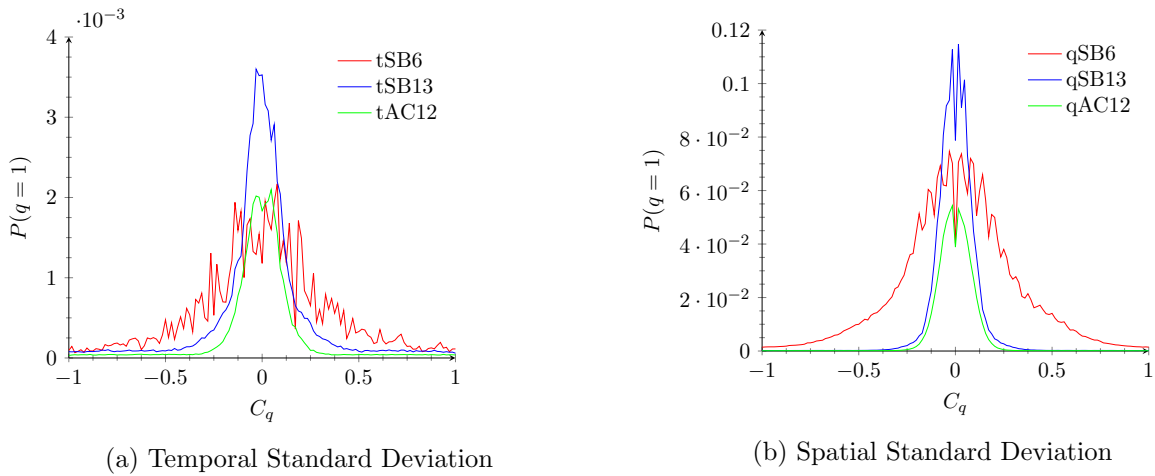

    \begin{subfigure}{0.49\textwidth}
	\scalebox{0.75}{
	    \begin{tikzpicture}
		\input{tab/QubitStDev}
		  \begin{axis} [
      axis lines=left,
      xlabel={$C_q$},
      ylabel={$P(q=1)$},
      ymin=0, ymax=0.004,
      xmin=-1, xmax=1,
      minor x tick num=3,
      minor y tick num=3,
      legend cell align=left,
      legend style={ draw=none, },
    ]
%    \addplot[red]     table[ x index=1, y index=2 ]   {};
    \addplot[red]     table[ x index=1, y index=2 ]   {\plotdata};
    \addplot[blue]    table[ x index=1, y index=4 ]   {\plotdata};
    \addplot[green]   table[ x index=1, y index=6 ]   {\plotdata};
    \legend{tSB6,tSB13,tAC12};
  \end{axis}
	    \end{tikzpicture}
	}
	\caption{Temporal Standard Deviation} \label{fig:qubitTstd}
    \end{subfigure}
    \hspace*{\fill} % separation between the subfigures
    \begin{subfigure}{0.49\textwidth}
	\scalebox{0.75}{
	    \begin{tikzpicture}
		\input{tab/QubitStDev}
		  \begin{axis} [
      axis lines=left,
      xlabel={$C_q$},
      ylabel={$P(q=1)$},
      ymin=0, ymax=0.12,
      xmin=-1, xmax=1,
      minor x tick num=3,
      minor y tick num=3,
      legend cell align=left,
      legend style={ draw=none, },
    ]
%    \addplot[red]     table[ x index=1, y index=2 ]   {};
    \addplot[red]     table[ x index=1, y index=3 ]   {\plotdata};
    \addplot[blue]    table[ x index=1, y index=5 ]   {\plotdata};
    \addplot[green]   table[ x index=1, y index=7 ]   {\plotdata};
    \legend{qSB6,qSB13,qAC12}
  \end{axis}
	    \end{tikzpicture}
	}
	\caption{Spatial Standard Deviation} \label{fig:qubitQstd}
    \end{subfigure}
    \caption{Plots of the variability of the probabilities of $q=1$ over (a)runs and (b)qubits. }
    \label{fig:qubitSTD}
\end{figure}

The effects of the environment can also be seen in figure \ref{fig:qubitSTD}. 
Figure \ref{fig:qubitTstd} is the standard deviation of the probabilities over time.  
Time here is represented by 10000 D-Wave samples. 
The samples were divided into 10 partitions and the standard deviation was computed across the partitions using the average probabilities over all samples and qubits for each partition. 
Figure \ref{fig:qubitQstd} is the standard deviation of the probabilities over space (or qubits). 
The average probability was computed for each qubit over all time partitions and the standard deviation was computed across the qubits of the D-Wave. 
Note that the peak temporal standard deviation ($1.13x10^{-4}$) is much lower than the peak spatial standard deviation ($1.45x10^{-3}$).  
This would indicate the properties of a qubit vary much less with time than amongst the qubits. 
Note also that the temporal STD of SB13 is higher than AC12 or SB6 and has a broader and noisier temporal STD than AC12. 
It can also be seen the the spatial STD of SB6 is broader and noisier than SB13 or AC12 while SB13's peak spatial STD is higher than either SB6 or AC12.  
Since SB6 is an older version than SB13 or AC12 and SB13 was an early test prototype of AC12 this is all quite understandable.  
The important point is that the quantum environmental isolation of the D-Wave is improving.

\subsection{Properties of a Qubit Chain}\label{sec:chainmetric}

Since the D-Wave architecture is based on a Chimera graph rather than a completely connected graph (clique), it is necessary to form virtual qubits out of sets of physical qubit. 
The idea is to configure the qubits and couplers to act as a single qubit that can be coupled to more qubits which are more widely dispersed across the D-Wave. 
The premise behind the virtual qubit (qubit chain) is that the physical qubits making up the virtual qubit should all return the same value. 
That is if one qubit of the group returns a value of one all the qubits of the group return values of one or if one qubit returns a value of zero all the qubits of the group return values of zero. 
This however is seldom the case, even if the value of the virtual qubit should be one (all the qubits of the chain should return one) some of the qubits may return a value of zero. 

We have performed experiments with groups of qubits formed from chains of qubits. 
Thirty chains were selected from the working qubits of AC12 for performing the measurement of stochastic behavior of virtual qubits. Each virtual qubit consisted of a chain of 12 physical qubits.  
For the measurements every coefficient of the qubits of the qubit chain are set to $C_q$ and every coupler coefficient of the qubit chain are set to $C_c$. 
All other coefficients are set to zero. No two qubit chains have any couplers in common. 
The probability of a virtual qubit to be one is calculated by counting all qubits that have a value of one divided by the number of qubits in the chain. 
For a given $C_q$ and $C_c$ the D-Wave is run 1000 times and the virtual qubit probability is averaged over all the runs and all virtual qubits (30) per run.

\begin{figure}
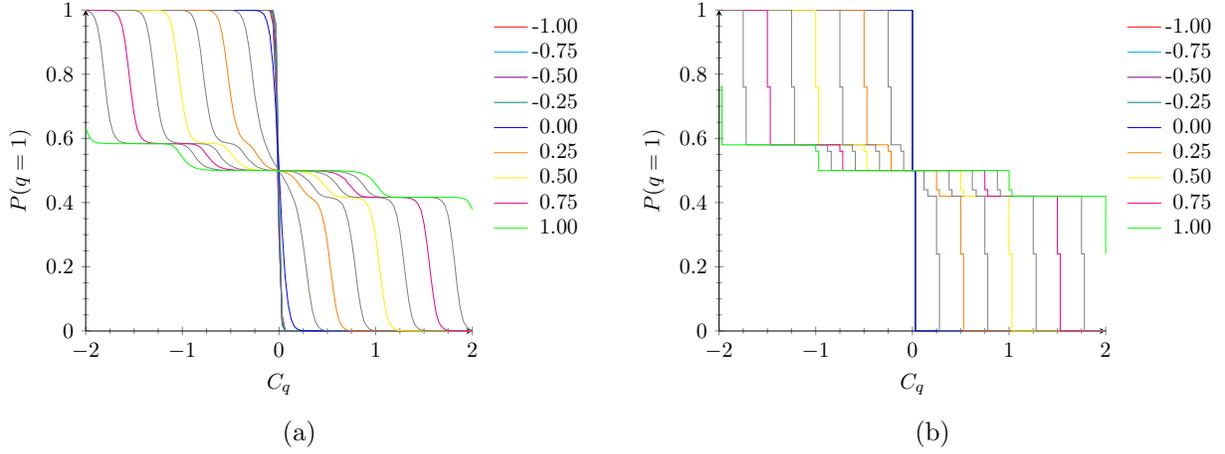

    \begin{subfigure}{0.49\textwidth}
	\scalebox{0.75}{
	    \begin{tikzpicture}
		\input{tab/ISING_dwave_129_017}
		  \begin{axis} [
      axis lines=left,
      xlabel={$C_q$},
      ylabel={$P(q=1)$},
      ymin=0, ymax=1,
      xmin=-2, xmax=2,
      minor x tick num=3,
      minor y tick num=3,
      legend cell align=right,
      legend style={ draw=none, at={(1.3,1.0)}, },
    ]
    \addplot[red]    table[ x index=1, y index=2,  ]                {\plotdata};
    \addplot[gray]   table[ x index=1, y index=3,  forget plot, ]   {\plotdata};
    \addplot[cyan]   table[ x index=1, y index=4,  ]                {\plotdata};
    \addplot[gray]   table[ x index=1, y index=5,  forget plot, ]   {\plotdata};
    \addplot[violet] table[ x index=1, y index=6,  ]                {\plotdata};
    \addplot[gray]   table[ x index=1, y index=7,  forget plot, ]   {\plotdata};
    \addplot[teal]   table[ x index=1, y index=8,  ]                {\plotdata};
    \addplot[gray]   table[ x index=1, y index=9,  forget plot, ]   {\plotdata};
    \addplot[blue]   table[ x index=1, y index=10, ]                {\plotdata};
    \addplot[gray]   table[ x index=1, y index=11, forget plot, ]   {\plotdata};
    \addplot[orange] table[ x index=1, y index=12, ]                {\plotdata};
    \addplot[gray]   table[ x index=1, y index=13, forget plot, ]   {\plotdata};
    \addplot[yellow] table[ x index=1, y index=14, ]                {\plotdata};
    \addplot[gray]   table[ x index=1, y index=15, forget plot, ]   {\plotdata};
    \addplot[magenta]table[ x index=1, y index=16, ]                {\plotdata};
    \addplot[gray]   table[ x index=1, y index=17, forget plot, ]   {\plotdata};
    \addplot[green]  table[ x index=1, y index=18, ]                {\plotdata};
    \legend{-1.00,-0.75,-0.50,-0.25,0.00,0.25,0.50,0.75,1.00}
  \end{axis}
	    \end{tikzpicture}
	}
	\caption{ } \label{fig:CHisingDcq}
    \end{subfigure}
    \hspace*{\fill} % separation between the subfigures
    \begin{subfigure}{0.49\textwidth}
	\scalebox{0.75}{
	    \begin{tikzpicture}
		\input{tab/ISING_theoretic_129_017}
		  \begin{axis} [
      axis lines=left,
      xlabel={$C_q$},
      ylabel={$P(q=1)$},
      ymin=0, ymax=1,
      xmin=-2, xmax=2,
      minor x tick num=3,
      minor y tick num=3,
      legend cell align=right,
      legend style={ draw=none, at={(1.3,1.0)}, },
    ]
    \addplot[red]    table[ x index=1, y index=2,  ]                {\plotdata};
    \addplot[gray]   table[ x index=1, y index=3,  forget plot, ]   {\plotdata};
    \addplot[cyan]   table[ x index=1, y index=4,  ]                {\plotdata};
    \addplot[gray]   table[ x index=1, y index=5,  forget plot, ]   {\plotdata};
    \addplot[violet] table[ x index=1, y index=6,  ]                {\plotdata};
    \addplot[gray]   table[ x index=1, y index=7,  forget plot, ]   {\plotdata};
    \addplot[teal]   table[ x index=1, y index=8,  ]                {\plotdata};
    \addplot[gray]   table[ x index=1, y index=9,  forget plot, ]   {\plotdata};
    \addplot[blue]   table[ x index=1, y index=10, ]                {\plotdata};
    \addplot[gray]   table[ x index=1, y index=11, forget plot, ]   {\plotdata};
    \addplot[orange] table[ x index=1, y index=12, ]                {\plotdata};
    \addplot[gray]   table[ x index=1, y index=13, forget plot, ]   {\plotdata};
    \addplot[yellow] table[ x index=1, y index=14, ]                {\plotdata};
    \addplot[gray]   table[ x index=1, y index=15, forget plot, ]   {\plotdata};
    \addplot[magenta]table[ x index=1, y index=16, ]                {\plotdata};
    \addplot[gray]   table[ x index=1, y index=17, forget plot, ]   {\plotdata};
    \addplot[green]  table[ x index=1, y index=18, ]                {\plotdata};
    \legend{-1.00,-0.75,-0.50,-0.25,0.00,0.25,0.50,0.75,1.00}
  \end{axis}
	    \end{tikzpicture}
	}
	\caption{ } \label{fig:CHisingTcq}
    \end{subfigure}
    \caption{Plots of $P(q=1)$ vs. $C_q$ using the Ising model for 12 qubit chains (a) on AC12 and (b) theoretically. (17 different values of $C_c$ were plotted.)  }
    \label{fig:CHisingCQ}
\end{figure}

\begin{figure}
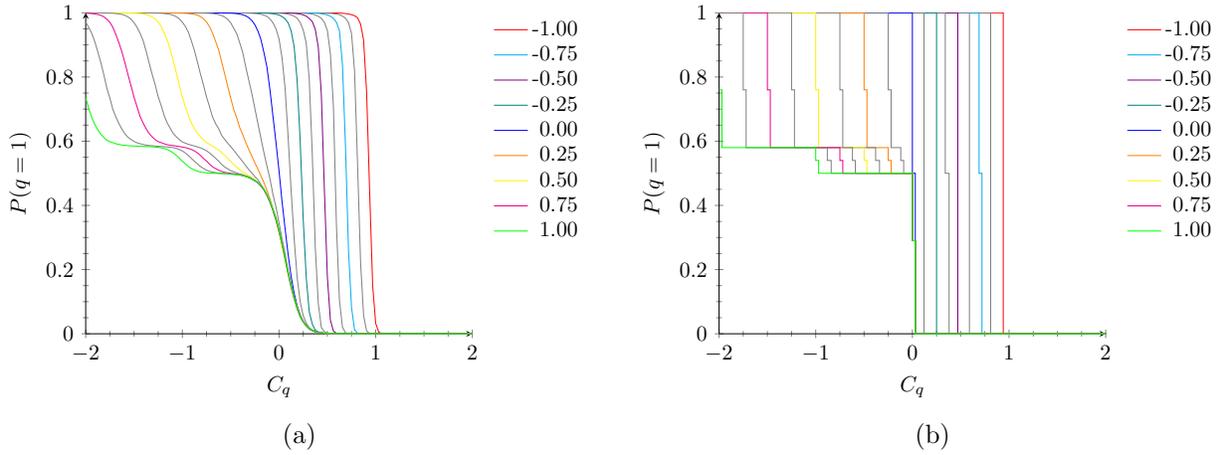

    \begin{subfigure}{0.49\textwidth}
	\scalebox{0.75}{
	    \begin{tikzpicture}
		\input{tab/QUBO_dwave_129_017}
		  \begin{axis} [
      axis lines=left,
      xlabel={$C_q$},
      ylabel={$P(q=1)$},
      ymin=0, ymax=1,
      xmin=-2, xmax=2,
      minor x tick num=3,
      minor y tick num=3,
      legend cell align=right,
      legend style={ draw=none, at={(1.3,1.0)}, },
    ]
    \addplot[red]    table[ x index=1, y index=2,  ]                {\plotdata};
    \addplot[gray]   table[ x index=1, y index=3,  forget plot, ]   {\plotdata};
    \addplot[cyan]   table[ x index=1, y index=4,  ]                {\plotdata};
    \addplot[gray]   table[ x index=1, y index=5,  forget plot, ]   {\plotdata};
    \addplot[violet] table[ x index=1, y index=6,  ]                {\plotdata};
    \addplot[gray]   table[ x index=1, y index=7,  forget plot, ]   {\plotdata};
    \addplot[teal]   table[ x index=1, y index=8,  ]                {\plotdata};
    \addplot[gray]   table[ x index=1, y index=9,  forget plot, ]   {\plotdata};
    \addplot[blue]   table[ x index=1, y index=10, ]                {\plotdata};
    \addplot[gray]   table[ x index=1, y index=11, forget plot, ]   {\plotdata};
    \addplot[orange] table[ x index=1, y index=12, ]                {\plotdata};
    \addplot[gray]   table[ x index=1, y index=13, forget plot, ]   {\plotdata};
    \addplot[yellow] table[ x index=1, y index=14, ]                {\plotdata};
    \addplot[gray]   table[ x index=1, y index=15, forget plot, ]   {\plotdata};
    \addplot[magenta]table[ x index=1, y index=16, ]                {\plotdata};
    \addplot[gray]   table[ x index=1, y index=17, forget plot, ]   {\plotdata};
    \addplot[green]  table[ x index=1, y index=18, ]                {\plotdata};
    \legend{-1.00,-0.75,-0.50,-0.25,0.00,0.25,0.50,0.75,1.00}
  \end{axis}
	    \end{tikzpicture}
	}
	\caption{ } \label{fig:CHquboDcq}
    \end{subfigure}
    \hspace*{\fill} % separation between the subfigures
    \begin{subfigure}{0.49\textwidth}
	\scalebox{0.75}{
	    \begin{tikzpicture}
		\input{tab/QUBO_theoretic_129_017}
		  \begin{axis} [
      axis lines=left,
      xlabel={$C_q$},
      ylabel={$P(q=1)$},
      ymin=0, ymax=1,
      xmin=-2, xmax=2,
      minor x tick num=3,
      minor y tick num=3,
      legend cell align=right,
      legend style={ draw=none, at={(1.3,1.0)}, },
    ]
    \addplot[red]    table[ x index=1, y index=2,  ]                {\plotdata};
    \addplot[gray]   table[ x index=1, y index=3,  forget plot, ]   {\plotdata};
    \addplot[cyan]   table[ x index=1, y index=4,  ]                {\plotdata};
    \addplot[gray]   table[ x index=1, y index=5,  forget plot, ]   {\plotdata};
    \addplot[violet] table[ x index=1, y index=6,  ]                {\plotdata};
    \addplot[gray]   table[ x index=1, y index=7,  forget plot, ]   {\plotdata};
    \addplot[teal]   table[ x index=1, y index=8,  ]                {\plotdata};
    \addplot[gray]   table[ x index=1, y index=9,  forget plot, ]   {\plotdata};
    \addplot[blue]   table[ x index=1, y index=10, ]                {\plotdata};
    \addplot[gray]   table[ x index=1, y index=11, forget plot, ]   {\plotdata};
    \addplot[orange] table[ x index=1, y index=12, ]                {\plotdata};
    \addplot[gray]   table[ x index=1, y index=13, forget plot, ]   {\plotdata};
    \addplot[yellow] table[ x index=1, y index=14, ]                {\plotdata};
    \addplot[gray]   table[ x index=1, y index=15, forget plot, ]   {\plotdata};
    \addplot[magenta]table[ x index=1, y index=16, ]                {\plotdata};
    \addplot[gray]   table[ x index=1, y index=17, forget plot, ]   {\plotdata};
    \addplot[green]  table[ x index=1, y index=18, ]                {\plotdata};
    \legend{-1.00,-0.75,-0.50,-0.25,0.00,0.25,0.50,0.75,1.00}
  \end{axis}
	    \end{tikzpicture}
	}
	\caption{ } \label{fig:CHquboTcq}
    \end{subfigure}
    \caption{Plots of $P(q=1)$ vs. $C_q$ using the QUBO model for 12 qubit chains (a) on AC12 and (b) theoretically. (17 different values of $C_c$ were plotted.)  }
    \label{fig:CHquboCQ}
\end{figure}

\begin{figure}
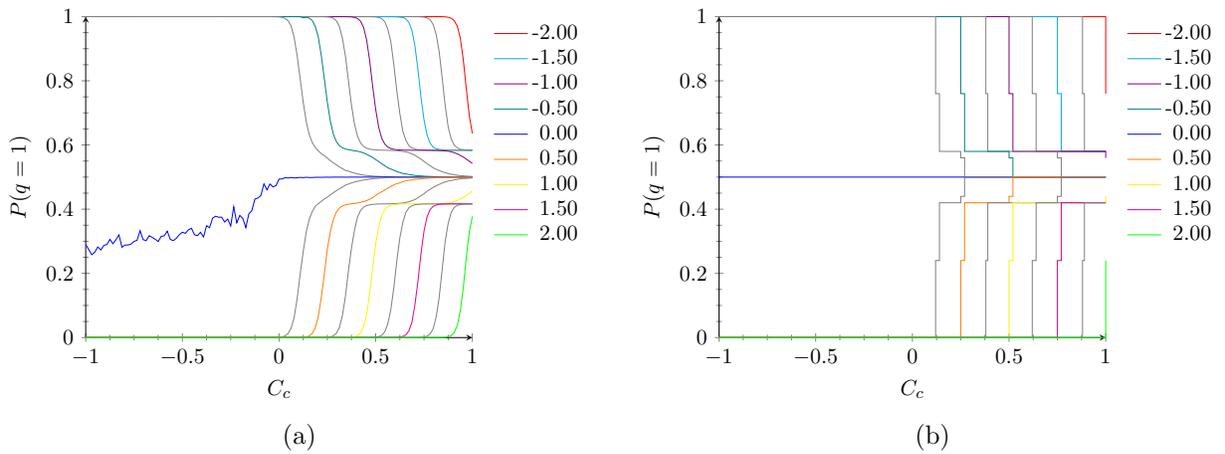

    \begin{subfigure}{0.49\textwidth}
	\scalebox{0.75}{
	    \begin{tikzpicture}
		\input{tab/ISING_dwave_017_129}
		  \begin{axis} [
      axis lines=left,
      xlabel={$C_c$},
      ylabel={$P(q=1)$},
      ymin=0, ymax=1,
      xmin=-1, xmax=1,
      minor x tick num=3,
      minor y tick num=3,
      legend cell align=right,
      legend style={ draw=none, at={(1.3,1.0)}, },
    ]
    \addplot[red]    table[ x index=1, y index=2,  ]                {\plotdata};
    \addplot[gray]   table[ x index=1, y index=3,  forget plot, ]   {\plotdata};
    \addplot[cyan]   table[ x index=1, y index=4,  ]                {\plotdata};
    \addplot[gray]   table[ x index=1, y index=5,  forget plot, ]   {\plotdata};
    \addplot[violet] table[ x index=1, y index=6,  ]                {\plotdata};
    \addplot[gray]   table[ x index=1, y index=7,  forget plot, ]   {\plotdata};
    \addplot[teal]   table[ x index=1, y index=8,  ]                {\plotdata};
    \addplot[gray]   table[ x index=1, y index=9,  forget plot, ]   {\plotdata};
    \addplot[blue]   table[ x index=1, y index=10, ]                {\plotdata};
    \addplot[gray]   table[ x index=1, y index=11, forget plot, ]   {\plotdata};
    \addplot[orange] table[ x index=1, y index=12, ]                {\plotdata};
    \addplot[gray]   table[ x index=1, y index=13, forget plot, ]   {\plotdata};
    \addplot[yellow] table[ x index=1, y index=14, ]                {\plotdata};
    \addplot[gray]   table[ x index=1, y index=15, forget plot, ]   {\plotdata};
    \addplot[magenta]table[ x index=1, y index=16, ]                {\plotdata};
    \addplot[gray]   table[ x index=1, y index=17, forget plot, ]   {\plotdata};
    \addplot[green]  table[ x index=1, y index=18, ]                {\plotdata};
    \legend{-2.00,-1.50,-1.00,-0.50,0.00,0.50,1.00,1.50,2.00}
  \end{axis}
	    \end{tikzpicture}
	}
	\caption{ } \label{fig:CHisingDqc}
    \end{subfigure}
    \hspace*{\fill} % separation between the subfigures
    \begin{subfigure}{0.49\textwidth}
	\scalebox{0.75}{
	    \begin{tikzpicture}
		\input{tab/ISING_theoretic_017_129}
		  \begin{axis} [
      axis lines=left,
      xlabel={$C_c$},
      ylabel={$P(q=1)$},
      ymin=0, ymax=1,
      xmin=-1, xmax=1,
      minor x tick num=3,
      minor y tick num=3,
      legend cell align=right,
      legend style={ draw=none, at={(1.3,1.0)}, },
    ]
    \addplot[red]    table[ x index=1, y index=2,  ]                {\plotdata};
    \addplot[gray]   table[ x index=1, y index=3,  forget plot, ]   {\plotdata};
    \addplot[cyan]   table[ x index=1, y index=4,  ]                {\plotdata};
    \addplot[gray]   table[ x index=1, y index=5,  forget plot, ]   {\plotdata};
    \addplot[violet] table[ x index=1, y index=6,  ]                {\plotdata};
    \addplot[gray]   table[ x index=1, y index=7,  forget plot, ]   {\plotdata};
    \addplot[teal]   table[ x index=1, y index=8,  ]                {\plotdata};
    \addplot[gray]   table[ x index=1, y index=9,  forget plot, ]   {\plotdata};
    \addplot[blue]   table[ x index=1, y index=10, ]                {\plotdata};
    \addplot[gray]   table[ x index=1, y index=11, forget plot, ]   {\plotdata};
    \addplot[orange] table[ x index=1, y index=12, ]                {\plotdata};
    \addplot[gray]   table[ x index=1, y index=13, forget plot, ]   {\plotdata};
    \addplot[yellow] table[ x index=1, y index=14, ]                {\plotdata};
    \addplot[gray]   table[ x index=1, y index=15, forget plot, ]   {\plotdata};
    \addplot[magenta]table[ x index=1, y index=16, ]                {\plotdata};
    \addplot[gray]   table[ x index=1, y index=17, forget plot, ]   {\plotdata};
    \addplot[green]  table[ x index=1, y index=18, ]                {\plotdata};
    \legend{-2.00,-1.50,-1.00,-0.50,0.00,0.50,1.00,1.50,2.00}
  \end{axis}
	    \end{tikzpicture}
	}
	\caption{ } \label{fig:CHisingTqc}
    \end{subfigure}
    \caption{Plots of $P(q=1)$ vs. $C_c$ using the Ising model for 12 qubit chains (a) on AC12 and (b) theoretically. (17 different values of $C_q$ were plotted.)  }
    \label{fig:CHisingQC}
\end{figure}

\begin{figure}
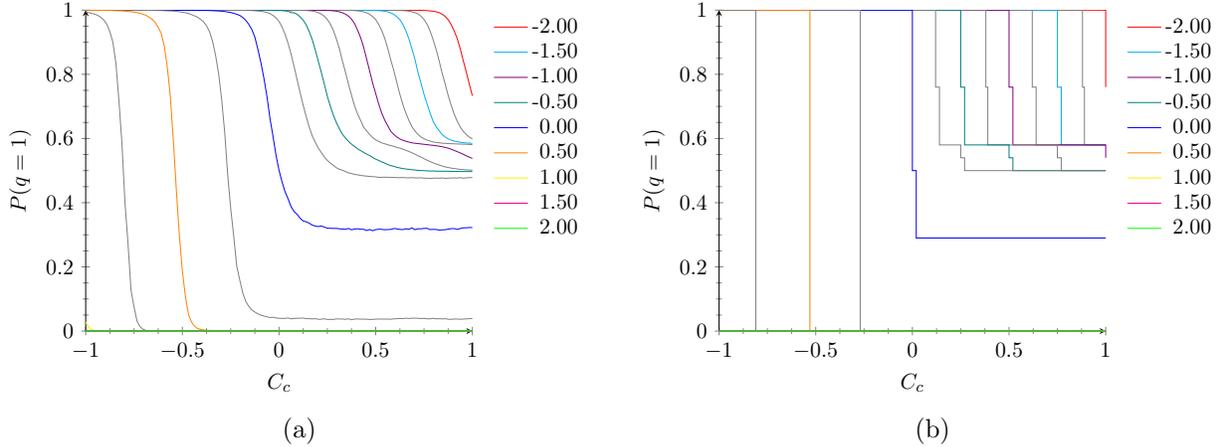

    \begin{subfigure}{0.49\textwidth}
	\scalebox{0.75}{
	    \begin{tikzpicture}
		\input{tab/QUBO_dwave_017_129}
		  \begin{axis} [
      axis lines=left,
      xlabel={$C_c$},
      ylabel={$P(q=1)$},
      ymin=0, ymax=1,
      xmin=-1, xmax=1,
      minor x tick num=3,
      minor y tick num=3,
      legend cell align=right,
      legend style={ draw=none, at={(1.3,1.0)}, },
    ]
    \addplot[red]    table[ x index=1, y index=2,  ]                {\plotdata};
    \addplot[gray]   table[ x index=1, y index=3,  forget plot, ]   {\plotdata};
    \addplot[cyan]   table[ x index=1, y index=4,  ]                {\plotdata};
    \addplot[gray]   table[ x index=1, y index=5,  forget plot, ]   {\plotdata};
    \addplot[violet] table[ x index=1, y index=6,  ]                {\plotdata};
    \addplot[gray]   table[ x index=1, y index=7,  forget plot, ]   {\plotdata};
    \addplot[teal]   table[ x index=1, y index=8,  ]                {\plotdata};
    \addplot[gray]   table[ x index=1, y index=9,  forget plot, ]   {\plotdata};
    \addplot[blue]   table[ x index=1, y index=10, ]                {\plotdata};
    \addplot[gray]   table[ x index=1, y index=11, forget plot, ]   {\plotdata};
    \addplot[orange] table[ x index=1, y index=12, ]                {\plotdata};
    \addplot[gray]   table[ x index=1, y index=13, forget plot, ]   {\plotdata};
    \addplot[yellow] table[ x index=1, y index=14, ]                {\plotdata};
    \addplot[gray]   table[ x index=1, y index=15, forget plot, ]   {\plotdata};
    \addplot[magenta]table[ x index=1, y index=16, ]                {\plotdata};
    \addplot[gray]   table[ x index=1, y index=17, forget plot, ]   {\plotdata};
    \addplot[green]  table[ x index=1, y index=18, ]                {\plotdata};
    \legend{-2.00,-1.50,-1.00,-0.50,0.00,0.50,1.00,1.50,2.00}
  \end{axis}
	    \end{tikzpicture}
	}
	\caption{ } \label{fig:CHquboDqc}
    \end{subfigure}
    \hspace*{\fill} % separation between the subfigures
    \begin{subfigure}{0.49\textwidth}
	\scalebox{0.75}{
	    \begin{tikzpicture}
		\input{tab/QUBO_theoretic_017_129}
		  \begin{axis} [
      axis lines=left,
      xlabel={$C_c$},
      ylabel={$P(q=1)$},
      ymin=0, ymax=1,
      xmin=-1, xmax=1,
      minor x tick num=3,
      minor y tick num=3,
      legend cell align=right,
      legend style={ draw=none, at={(1.3,1.0)}, },
    ]
    \addplot[red]    table[ x index=1, y index=2,  ]                {\plotdata};
    \addplot[gray]   table[ x index=1, y index=3,  forget plot, ]   {\plotdata};
    \addplot[cyan]   table[ x index=1, y index=4,  ]                {\plotdata};
    \addplot[gray]   table[ x index=1, y index=5,  forget plot, ]   {\plotdata};
    \addplot[violet] table[ x index=1, y index=6,  ]                {\plotdata};
    \addplot[gray]   table[ x index=1, y index=7,  forget plot, ]   {\plotdata};
    \addplot[teal]   table[ x index=1, y index=8,  ]                {\plotdata};
    \addplot[gray]   table[ x index=1, y index=9,  forget plot, ]   {\plotdata};
    \addplot[blue]   table[ x index=1, y index=10, ]                {\plotdata};
    \addplot[gray]   table[ x index=1, y index=11, forget plot, ]   {\plotdata};
    \addplot[orange] table[ x index=1, y index=12, ]                {\plotdata};
    \addplot[gray]   table[ x index=1, y index=13, forget plot, ]   {\plotdata};
    \addplot[yellow] table[ x index=1, y index=14, ]                {\plotdata};
    \addplot[gray]   table[ x index=1, y index=15, forget plot, ]   {\plotdata};
    \addplot[magenta]table[ x index=1, y index=16, ]                {\plotdata};
    \addplot[gray]   table[ x index=1, y index=17, forget plot, ]   {\plotdata};
    \addplot[green]  table[ x index=1, y index=18, ]                {\plotdata};
    \legend{-2.00,-1.50,-1.00,-0.50,0.00,0.50,1.00,1.50,2.00}
  \end{axis}
	    \end{tikzpicture}
	}
	\caption{ } \label{fig:CHquboTqc}
    \end{subfigure}
    \caption{Plots of $P(q=1)$ vs. $C_c$ using the QUBO model for 12 qubit chains (a) on AC12 and (b) theoretically. (17 different values of $C_q$ were plotted.)  }
    \label{fig:CHquboQC}
\end{figure}

Figures \ref{fig:CHisingCQ}-\ref{fig:CHquboQC} are plots of families of characteristic curves that show the stochastic properties of 12 qubit virtual qubits. 
Ensembles were run with a fixed value of $C_q$ between 2 and -2 and a fixed value of $C_c$ between 1 and -1. 
Each ensemble was an average over 1000 D-Wave runs of 30 virtual qubits (qubit chains). 
The ensembles were run using the Ising model (where qubits can have a value of -1 or 1) and the QUBO model (where qubits can have a value of 0 or 1). 
The D-Wave is based on the Ising model but can be coerced into running as a QUBO model through algebraic manipulation of the qubit and coupler coefficients.
Figure \ref{fig:CHisingCQ} is based on the Ising model. 
Each line in the plot is for a specific value of $C_c$.
There are 17 plots for different values of $C_c$ in the range [-1,1]. 
Figure \ref{fig:CHquboCQ} is similar to \ref{fig:CHisingCQ} using the QUBO model instead of Ising. 
Figure \ref{fig:CHisingQC} is based on the Ising model. 
Each line in the plot is a for a specific value of $C_q$. 
There are 17 plots with different values of $C_q$ in the range [-2,2]. 
Figures \ref{fig:CHquboQC} is similar to \ref{fig:CHisingQC} using the QUBO model instead of Ising.

Figures \ref{fig:CHisingDcq}, \ref{fig:CHquboDcq}, \ref{fig:CHisingDqc}, and \ref{fig:CHquboDqc} are plotted from measurement taken from AC12.
While figures \ref{fig:CHisingTcq}, \ref{fig:CHquboTcq}, \ref{fig:CHisingTqc}, and \ref{fig:CHquboTqc} are plots of what a theoretically perfect machine would produce if given the same coefficients that were given the AC12. 
This theoretically perfect machine($T_e=0$) is based on a machine which would always return a set of values of qubits which minimize the objective function, equation \ref{eq:obfunc}, thus, it represents the stochastic behavior of an errorless/noiseless D-Wave.  
These theoretic predictions were calculated by computing the value of equation \ref{eq:obfunc} for all possible cases of 12 qubits (4096) for each pair of values of $C_q$ and $C_c$. 
The resultant theoretical probability($P(q=1)$) for each value pair is the average probability over all global minimum states. 
Fundamentally the D-Wave plots are very similar to the theoretical result, but differ in the sharpness of the curves as was the case in figure \ref{fig:qubitprob}. 
This will effect how probable it is that the resultant value of the objective function will be the global minimum or how close that value is to the global minimum. 
Note that qubit chains only behave properly as describe previously while using the QUBO model and $C_q$ is within the range [0,1] and $C_c$ is within the range [-1,0].

\subsection{Properties of a Qubit Cell}\label{sec:cellmetric}

\begin{figure}
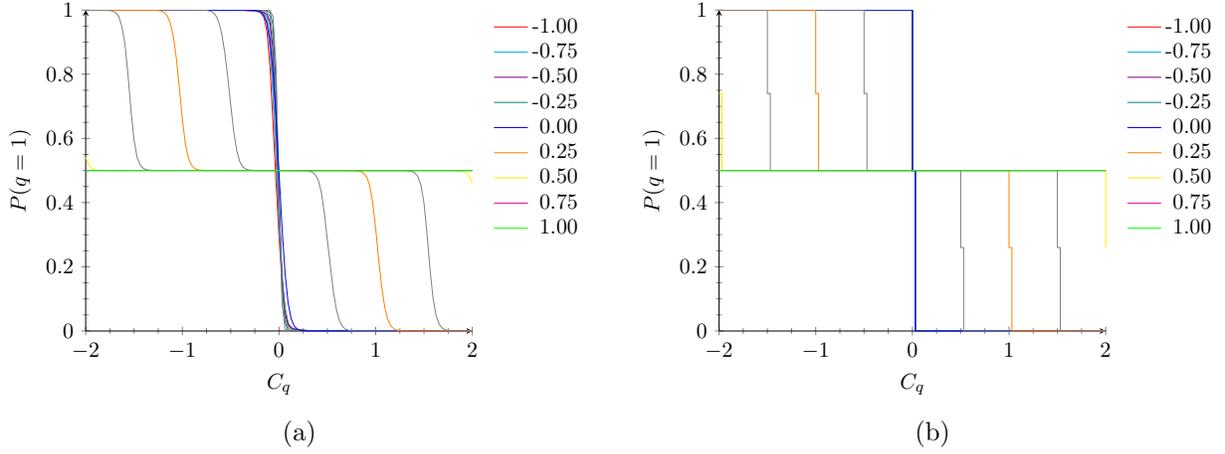

    \begin{subfigure}{0.49\textwidth}
	\scalebox{0.75}{
	    \begin{tikzpicture}
		\input{tab/cell_ISING_dwave_129_017}
		  \begin{axis} [
      axis lines=left,
      xlabel={$C_q$},
      ylabel={$P(q=1)$},
      ymin=0, ymax=1,
      xmin=-2, xmax=2,
      minor x tick num=3,
      minor y tick num=3,
      legend cell align=right,
      legend style={ draw=none, at={(1.3,1.0)}, },
    ]
    \addplot[red]    table[ x index=1, y index=2,  ]                {\plotdata};
    \addplot[gray]   table[ x index=1, y index=3,  forget plot, ]   {\plotdata};
    \addplot[cyan]   table[ x index=1, y index=4,  ]                {\plotdata};
    \addplot[gray]   table[ x index=1, y index=5,  forget plot, ]   {\plotdata};
    \addplot[violet] table[ x index=1, y index=6,  ]                {\plotdata};
    \addplot[gray]   table[ x index=1, y index=7,  forget plot, ]   {\plotdata};
    \addplot[teal]   table[ x index=1, y index=8,  ]                {\plotdata};
    \addplot[gray]   table[ x index=1, y index=9,  forget plot, ]   {\plotdata};
    \addplot[blue]   table[ x index=1, y index=10, ]                {\plotdata};
    \addplot[gray]   table[ x index=1, y index=11, forget plot, ]   {\plotdata};
    \addplot[orange] table[ x index=1, y index=12, ]                {\plotdata};
    \addplot[gray]   table[ x index=1, y index=13, forget plot, ]   {\plotdata};
    \addplot[yellow] table[ x index=1, y index=14, ]                {\plotdata};
    \addplot[gray]   table[ x index=1, y index=15, forget plot, ]   {\plotdata};
    \addplot[magenta]table[ x index=1, y index=16, ]                {\plotdata};
    \addplot[gray]   table[ x index=1, y index=17, forget plot, ]   {\plotdata};
    \addplot[green]  table[ x index=1, y index=18, ]                {\plotdata};
    \legend{-1.00,-0.75,-0.50,-0.25,0.00,0.25,0.50,0.75,1.00}
  \end{axis}
	    \end{tikzpicture}
	}
	\caption{ } \label{fig:CEisingDcq}
    \end{subfigure}
    \hspace*{\fill} % separation between the subfigures
    \begin{subfigure}{0.49\textwidth}
	\scalebox{0.75}{
	    \begin{tikzpicture}
		\input{tab/cell_ISING_theoretic_129_017}
		  \begin{axis} [
      axis lines=left,
      xlabel={$C_q$},
      ylabel={$P(q=1)$},
      ymin=0, ymax=1,
      xmin=-2, xmax=2,
      minor x tick num=3,
      minor y tick num=3,
      legend cell align=right,
      legend style={ draw=none, at={(1.3,1.0)}, },
    ]
    \addplot[red]    table[ x index=1, y index=2,  ]                {\plotdata};
    \addplot[gray]   table[ x index=1, y index=3,  forget plot, ]   {\plotdata};
    \addplot[cyan]   table[ x index=1, y index=4,  ]                {\plotdata};
    \addplot[gray]   table[ x index=1, y index=5,  forget plot, ]   {\plotdata};
    \addplot[violet] table[ x index=1, y index=6,  ]                {\plotdata};
    \addplot[gray]   table[ x index=1, y index=7,  forget plot, ]   {\plotdata};
    \addplot[teal]   table[ x index=1, y index=8,  ]                {\plotdata};
    \addplot[gray]   table[ x index=1, y index=9,  forget plot, ]   {\plotdata};
    \addplot[blue]   table[ x index=1, y index=10, ]                {\plotdata};
    \addplot[gray]   table[ x index=1, y index=11, forget plot, ]   {\plotdata};
    \addplot[orange] table[ x index=1, y index=12, ]                {\plotdata};
    \addplot[gray]   table[ x index=1, y index=13, forget plot, ]   {\plotdata};
    \addplot[yellow] table[ x index=1, y index=14, ]                {\plotdata};
    \addplot[gray]   table[ x index=1, y index=15, forget plot, ]   {\plotdata};
    \addplot[magenta]table[ x index=1, y index=16, ]                {\plotdata};
    \addplot[gray]   table[ x index=1, y index=17, forget plot, ]   {\plotdata};
    \addplot[green]  table[ x index=1, y index=18, ]                {\plotdata};
    \legend{-1.00,-0.75,-0.50,-0.25,0.00,0.25,0.50,0.75,1.00}
  \end{axis}
	    \end{tikzpicture}
	}
	\caption{ } \label{fig:CEisingTcq}
    \end{subfigure}
    \caption{Plots of $P(q=1)$ vs. $C_q$ using the Ising model for 8 qubit cells (a) on AC12 and (b) theoretically. (17 different values of $C_c$ were plotted.)  }
    \label{fig:CEisingCQ}
\end{figure}

The D-Wave 2X consists of 12x12 or 144 8 qubit cells. Each cell consists of a 4x4 qubit bipartite graph. 
The AC12 has 108 such complete cells, having 8 working qubits. 
Though the qubit cell is not as significant as the qubit chain in algorithm design for the D-Wave, it still presents interesting conformation of the properties of the D-Wave. 
Figures \ref{fig:CEisingDcq}, \ref{fig:CEquboDcq}, \ref{fig:CEisingDqc}, and \ref{fig:CEquboDqc} corespond to figures \ref{fig:CHisingDcq}, \ref{fig:CHquboDcq}, \ref{fig:CHisingDqc}, and \ref{fig:CHquboDqc}, using cells of qubits rather than chains.  Figures \ref{fig:CEisingTcq}, \ref{fig:CEquboTcq}, \ref{fig:CEisingTqc}, and \ref{fig:CEquboTqc} are the theoretical versions of cell behavior coresponding to figures \ref{fig:CHisingTcq}, \ref{fig:CHquboTcq}, \ref{fig:CHisingTqc}, and \ref{fig:CHquboTqc} for chains. 
\begin{figure}
    \begin{subfigure}{0.49\textwidth}
	\scalebox{0.75}{
	    \begin{tikzpicture}
		\input{tab/cell_QUBO_dwave_129_017}
		  \begin{axis} [
      axis lines=left,
      xlabel={$C_q$},
      ylabel={$P(q=1)$},
      ymin=0, ymax=1,
      xmin=-2, xmax=2,
      minor x tick num=3,
      minor y tick num=3,
      legend cell align=right,
      legend style={ draw=none, at={(1.3,1.0)}, },
    ]
    \addplot[red]    table[ x index=1, y index=2,  ]                {\plotdata};
    \addplot[gray]   table[ x index=1, y index=3,  forget plot, ]   {\plotdata};
    \addplot[cyan]   table[ x index=1, y index=4,  ]                {\plotdata};
    \addplot[gray]   table[ x index=1, y index=5,  forget plot, ]   {\plotdata};
    \addplot[violet] table[ x index=1, y index=6,  ]                {\plotdata};
    \addplot[gray]   table[ x index=1, y index=7,  forget plot, ]   {\plotdata};
    \addplot[teal]   table[ x index=1, y index=8,  ]                {\plotdata};
    \addplot[gray]   table[ x index=1, y index=9,  forget plot, ]   {\plotdata};
    \addplot[blue]   table[ x index=1, y index=10, ]                {\plotdata};
    \addplot[gray]   table[ x index=1, y index=11, forget plot, ]   {\plotdata};
    \addplot[orange] table[ x index=1, y index=12, ]                {\plotdata};
    \addplot[gray]   table[ x index=1, y index=13, forget plot, ]   {\plotdata};
    \addplot[yellow] table[ x index=1, y index=14, ]                {\plotdata};
    \addplot[gray]   table[ x index=1, y index=15, forget plot, ]   {\plotdata};
    \addplot[magenta]table[ x index=1, y index=16, ]                {\plotdata};
    \addplot[gray]   table[ x index=1, y index=17, forget plot, ]   {\plotdata};
    \addplot[green]  table[ x index=1, y index=18, ]                {\plotdata};
    \legend{-1.00,-0.75,-0.50,-0.25,0.00,0.25,0.50,0.75,1.00}
  \end{axis}
	    \end{tikzpicture}
	}
	\caption{ } \label{fig:CEquboDcq}
    \end{subfigure}
    \hspace*{\fill} % separation between the subfigures
    \begin{subfigure}{0.49\textwidth}
	\scalebox{0.75}{
	    \begin{tikzpicture}
		\input{tab/cell_QUBO_theoretic_129_017}
		  \begin{axis} [
      axis lines=left,
      xlabel={$C_q$},
      ylabel={$P(q=1)$},
      ymin=0, ymax=1,
      xmin=-2, xmax=2,
      minor x tick num=3,
      minor y tick num=3,
      legend cell align=right,
      legend style={ draw=none, at={(1.3,1.0)}, },
    ]
    \addplot[red]    table[ x index=1, y index=2,  ]                {\plotdata};
    \addplot[gray]   table[ x index=1, y index=3,  forget plot, ]   {\plotdata};
    \addplot[cyan]   table[ x index=1, y index=4,  ]                {\plotdata};
    \addplot[gray]   table[ x index=1, y index=5,  forget plot, ]   {\plotdata};
    \addplot[violet] table[ x index=1, y index=6,  ]                {\plotdata};
    \addplot[gray]   table[ x index=1, y index=7,  forget plot, ]   {\plotdata};
    \addplot[teal]   table[ x index=1, y index=8,  ]                {\plotdata};
    \addplot[gray]   table[ x index=1, y index=9,  forget plot, ]   {\plotdata};
    \addplot[blue]   table[ x index=1, y index=10, ]                {\plotdata};
    \addplot[gray]   table[ x index=1, y index=11, forget plot, ]   {\plotdata};
    \addplot[orange] table[ x index=1, y index=12, ]                {\plotdata};
    \addplot[gray]   table[ x index=1, y index=13, forget plot, ]   {\plotdata};
    \addplot[yellow] table[ x index=1, y index=14, ]                {\plotdata};
    \addplot[gray]   table[ x index=1, y index=15, forget plot, ]   {\plotdata};
    \addplot[magenta]table[ x index=1, y index=16, ]                {\plotdata};
    \addplot[gray]   table[ x index=1, y index=17, forget plot, ]   {\plotdata};
    \addplot[green]  table[ x index=1, y index=18, ]                {\plotdata};
    \legend{-1.00,-0.75,-0.50,-0.25,0.00,0.25,0.50,0.75,1.00}
  \end{axis}
	    \end{tikzpicture}
	}
	\caption{ } \label{fig:CEquboTcq}
    \end{subfigure}
    \caption{Plots of $P(q=1)$ vs. $C_q$ using the QUBO model for 8 qubit cells (a) on AC12 and (b) theoretically. (17 different values of $C_c$ were plotted.)  }
    \label{fig:CEquboCQ}
\end{figure}
\begin{figure}
    \begin{subfigure}{0.49\textwidth}
	\scalebox{0.75}{
	    \begin{tikzpicture}
		\input{tab/cell_ISING_dwave_017_129}
		  \begin{axis} [
      axis lines=left,
      xlabel={$C_c$},
      ylabel={$P(q=1)$},
      ymin=0, ymax=1,
      xmin=-1, xmax=1,
      minor x tick num=3,
      minor y tick num=3,
      legend cell align=right,
      legend style={ draw=none, at={(1.3,1.0)}, },
    ]
    \addplot[red]    table[ x index=1, y index=2,  ]                {\plotdata};
    \addplot[gray]   table[ x index=1, y index=3,  forget plot, ]   {\plotdata};
    \addplot[cyan]   table[ x index=1, y index=4,  ]                {\plotdata};
    \addplot[gray]   table[ x index=1, y index=5,  forget plot, ]   {\plotdata};
    \addplot[violet] table[ x index=1, y index=6,  ]                {\plotdata};
    \addplot[gray]   table[ x index=1, y index=7,  forget plot, ]   {\plotdata};
    \addplot[teal]   table[ x index=1, y index=8,  ]                {\plotdata};
    \addplot[gray]   table[ x index=1, y index=9,  forget plot, ]   {\plotdata};
    \addplot[blue]   table[ x index=1, y index=10, ]                {\plotdata};
    \addplot[gray]   table[ x index=1, y index=11, forget plot, ]   {\plotdata};
    \addplot[orange] table[ x index=1, y index=12, ]                {\plotdata};
    \addplot[gray]   table[ x index=1, y index=13, forget plot, ]   {\plotdata};
    \addplot[yellow] table[ x index=1, y index=14, ]                {\plotdata};
    \addplot[gray]   table[ x index=1, y index=15, forget plot, ]   {\plotdata};
    \addplot[magenta]table[ x index=1, y index=16, ]                {\plotdata};
    \addplot[gray]   table[ x index=1, y index=17, forget plot, ]   {\plotdata};
    \addplot[green]  table[ x index=1, y index=18, ]                {\plotdata};
    \legend{-2.00,-1.50,-1.00,-0.50,0.00,0.50,1.00,1.50,2.00}
  \end{axis}
	    \end{tikzpicture}
	}
	\caption{ } \label{fig:CEisingDqc}
    \end{subfigure}
    \hspace*{\fill} % separation between the subfigures
    \begin{subfigure}{0.49\textwidth}
	\scalebox{0.75}{
	    \begin{tikzpicture}
		\input{tab/cell_ISING_theoretic_017_129}
		  \begin{axis} [
      axis lines=left,
      xlabel={$C_c$},
      ylabel={$P(q=1)$},
      ymin=0, ymax=1,
      xmin=-1, xmax=1,
      minor x tick num=3,
      minor y tick num=3,
      legend cell align=right,
      legend style={ draw=none, at={(1.3,1.0)}, },
    ]
    \addplot[red]    table[ x index=1, y index=2,  ]                {\plotdata};
    \addplot[gray]   table[ x index=1, y index=3,  forget plot, ]   {\plotdata};
    \addplot[cyan]   table[ x index=1, y index=4,  ]                {\plotdata};
    \addplot[gray]   table[ x index=1, y index=5,  forget plot, ]   {\plotdata};
    \addplot[violet] table[ x index=1, y index=6,  ]                {\plotdata};
    \addplot[gray]   table[ x index=1, y index=7,  forget plot, ]   {\plotdata};
    \addplot[teal]   table[ x index=1, y index=8,  ]                {\plotdata};
    \addplot[gray]   table[ x index=1, y index=9,  forget plot, ]   {\plotdata};
    \addplot[blue]   table[ x index=1, y index=10, ]                {\plotdata};
    \addplot[gray]   table[ x index=1, y index=11, forget plot, ]   {\plotdata};
    \addplot[orange] table[ x index=1, y index=12, ]                {\plotdata};
    \addplot[gray]   table[ x index=1, y index=13, forget plot, ]   {\plotdata};
    \addplot[yellow] table[ x index=1, y index=14, ]                {\plotdata};
    \addplot[gray]   table[ x index=1, y index=15, forget plot, ]   {\plotdata};
    \addplot[magenta]table[ x index=1, y index=16, ]                {\plotdata};
    \addplot[gray]   table[ x index=1, y index=17, forget plot, ]   {\plotdata};
    \addplot[green]  table[ x index=1, y index=18, ]                {\plotdata};
    \legend{-2.00,-1.50,-1.00,-0.50,0.00,0.50,1.00,1.50,2.00}
  \end{axis}
	    \end{tikzpicture}
	}
	\caption{ } \label{fig:CEisingTqc}
    \end{subfigure}
    \caption{Plots of $P(q=1)$ vs. $C_c$ using the Ising model for 8 qubit cells (a) on AC12 and (b) theoretically. (17 different values of $C_q$ were plotted.)  }
    \label{fig:CEisingQC}
\end{figure}
\begin{figure}
    \begin{subfigure}{0.49\textwidth}
	\scalebox{0.75}{
	    \begin{tikzpicture}
		\input{tab/cell_QUBO_dwave_017_129}
		  \begin{axis} [
      axis lines=left,
      xlabel={$C_c$},
      ylabel={$P(q=1)$},
      ymin=0, ymax=1,
      xmin=-1, xmax=1,
      minor x tick num=3,
      minor y tick num=3,
      legend cell align=right,
      legend style={ draw=none, at={(1.3,1.0)}, },
    ]
    \addplot[red]    table[ x index=1, y index=2,  ]                {\plotdata};
    \addplot[gray]   table[ x index=1, y index=3,  forget plot, ]   {\plotdata};
    \addplot[cyan]   table[ x index=1, y index=4,  ]                {\plotdata};
    \addplot[gray]   table[ x index=1, y index=5,  forget plot, ]   {\plotdata};
    \addplot[violet] table[ x index=1, y index=6,  ]                {\plotdata};
    \addplot[gray]   table[ x index=1, y index=7,  forget plot, ]   {\plotdata};
    \addplot[teal]   table[ x index=1, y index=8,  ]                {\plotdata};
    \addplot[gray]   table[ x index=1, y index=9,  forget plot, ]   {\plotdata};
    \addplot[blue]   table[ x index=1, y index=10, ]                {\plotdata};
    \addplot[gray]   table[ x index=1, y index=11, forget plot, ]   {\plotdata};
    \addplot[orange] table[ x index=1, y index=12, ]                {\plotdata};
    \addplot[gray]   table[ x index=1, y index=13, forget plot, ]   {\plotdata};
    \addplot[yellow] table[ x index=1, y index=14, ]                {\plotdata};
    \addplot[gray]   table[ x index=1, y index=15, forget plot, ]   {\plotdata};
    \addplot[magenta]table[ x index=1, y index=16, ]                {\plotdata};
    \addplot[gray]   table[ x index=1, y index=17, forget plot, ]   {\plotdata};
    \addplot[green]  table[ x index=1, y index=18, ]                {\plotdata};
    \legend{-2.00,-1.50,-1.00,-0.50,0.00,0.50,1.00,1.50,2.00}
  \end{axis}
	    \end{tikzpicture}
	}
	\caption{ } \label{fig:CEquboDqc}
    \end{subfigure}
    \hspace*{\fill} % separation between the subfigures
    \begin{subfigure}{0.49\textwidth}
	\scalebox{0.75}{
	    \begin{tikzpicture}
		\input{tab/cell_QUBO_theoretic_017_129}
		  \begin{axis} [
      axis lines=left,
      xlabel={$C_c$},
      ylabel={$P(q=1)$},
      ymin=0, ymax=1,
      xmin=-1, xmax=1,
      minor x tick num=3,
      minor y tick num=3,
      legend cell align=right,
      legend style={ draw=none, at={(1.3,1.0)}, },
    ]
    \addplot[red]    table[ x index=1, y index=2,  ]                {\plotdata};
    \addplot[gray]   table[ x index=1, y index=3,  forget plot, ]   {\plotdata};
    \addplot[cyan]   table[ x index=1, y index=4,  ]                {\plotdata};
    \addplot[gray]   table[ x index=1, y index=5,  forget plot, ]   {\plotdata};
    \addplot[violet] table[ x index=1, y index=6,  ]                {\plotdata};
    \addplot[gray]   table[ x index=1, y index=7,  forget plot, ]   {\plotdata};
    \addplot[teal]   table[ x index=1, y index=8,  ]                {\plotdata};
    \addplot[gray]   table[ x index=1, y index=9,  forget plot, ]   {\plotdata};
    \addplot[blue]   table[ x index=1, y index=10, ]                {\plotdata};
    \addplot[gray]   table[ x index=1, y index=11, forget plot, ]   {\plotdata};
    \addplot[orange] table[ x index=1, y index=12, ]                {\plotdata};
    \addplot[gray]   table[ x index=1, y index=13, forget plot, ]   {\plotdata};
    \addplot[yellow] table[ x index=1, y index=14, ]                {\plotdata};
    \addplot[gray]   table[ x index=1, y index=15, forget plot, ]   {\plotdata};
    \addplot[magenta]table[ x index=1, y index=16, ]                {\plotdata};
    \addplot[gray]   table[ x index=1, y index=17, forget plot, ]   {\plotdata};
    \addplot[green]  table[ x index=1, y index=18, ]                {\plotdata};
    \legend{-2.00,-1.50,-1.00,-0.50,0.00,0.50,1.00,1.50,2.00}
  \end{axis}
	    \end{tikzpicture}
	}
	\caption{ } \label{fig:CEquboTqc}
    \end{subfigure}
    \caption{Plots of $P(q=1)$ vs. $C_c$ using the QUBO model for 8 qubit cells (a) on AC12 and (b) theoretically. (17 different values of $C_q$ were plotted.)  }
    \label{fig:CEquboQC}
\end{figure}

\section{Conclusion}\label{sec:conclusion}
The purpose of this paper was to show the properties of the D-Wave that facilitate application development.  
In the process it was necessary to show what are the stable properties as well as the properties for which there is need for improvement.  
When starting this paper it appeared that the D-Wave was not performing as it was portrayed or at least as well as expected, that is inexplicably unpredictable and erratic. 
This begged for an explanation as to why this seemed to be the case.  
The first step was to make simple measurements of a simple configuration, how does a qubit behave. 
When stochastic measurements of the probabilities of a qubit, $P(q=1)$, given different coefficients (figure \ref{fig:qubitprob}) was made, it was clear that the qubit was behaving very predictably. 
But how well was it contributing to the minimization of the objective function (equation \ref{eq:obfunc})? 
Assuming that the behavior of the qubit needs to either match or be near the theoretically perfect behavior, figure \ref{fig:qubitprob} show that AC12 is much closer to the theoretically perfect behavior than SB6.
This paper has not studied if the AC12 behavior is good enough.
Next, stochastic measurements of qubit chains, needed to implement virtual qubits, were performed. Figures \ref{fig:CHisingCQ} thru \ref{fig:CHquboQC} present that study. 
It was thought that chains longer than a half a dozen qubits do not give very good results.  
The plots of qubits chains in this paper indicate this problem.  
The sharpness of the theoretical results in contrast to the D-Wave measurements would indicate why these qubit chains did not perform as desired. 
The simple conclusion is that the D-Wave qubit properties need to be improved by improving it's quantum environment isolation. 
In time this will happen as with all new technologies. 
In the present it is necessary for there to be work done in discovering more robust algorithms, that will adjust to the current and future frailties of this type of computational architecture.
Classical ones that can utilize what is returned by the D-Wave to enhance the classical solution either in speed or accuracy, a hybrid solution integrating classical algorithms with quantum ones.

\section*{Acknowledgement}

The author would like to thank Michael Little and Marjorie Cole of the NASA Advanced Information Systems 
Technology Office for their continued support for this research effort under grant NNX15AK58G and to the 
NASA Ames Research Center for providing access to the D-Wave quantum annealing computer. 
In addition, the author thanks the NSF funded Center for Hybrid Multicore Productivity Research and 
D-Wave Systems for their support and access to their computational resources.

\bibliography{QubitChains}{}

\newpage
\appendix
\title {\LARGE ADDENDUM}
\date{}
\maketitle
\section{Characteristics of Qubits and Qubit Chains Using Voting as the Metric}\label{sec:vote}

The metric in this paper for characterizing qubit chains/cells is
the average probability that a qubit in a chain/cell has a value of one. 
The metric in this addendum for characterizing qubit chains/cells is voting.
The voting metric is defined as the probability that the majority of qubits of a chain/cell are one.
For an even number of qubits if half of the qubits have a value of one the chain/cell wins the vote.
The following plots give the probability that a chain/cell wins the vote given a specific value of $C_q$ and $C_c$.
Each Figure contains the actual result from the D-Wave 2X (a) and the theoretical result (b) for each case. 
The cases show the Ising and the QUBO model, coupler weights and qubit biases, and qubit chains (virtual qubits) 
and qubit cells. 
 
Note that the plots of the voting results presented in this addendum are much less complex than the plots 
of results presented in the main body of the paper.
The theoretical plots as in the paper have sharper features than the D-Wave plots, yet they are structurally simmilar.
It appears that voting may be a better way of determining if a chain/cell is true (1) or false (-1 or 0), 
rather than simple probabilities or that all the qubits of a chain/cell be required to 
be true or false for the chain/cell to be treated as true or false.

\begin{figure}[H]
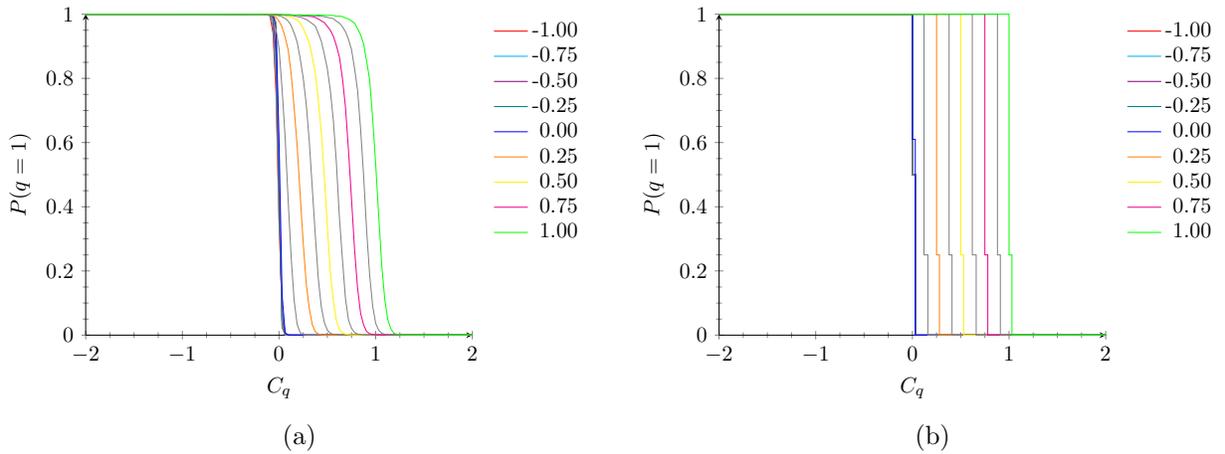

    \begin{subfigure}{0.49\textwidth}
	\scalebox{0.75}{
	    \begin{tikzpicture}
		\input{tab/chain_ISING_dwave_vote_129_017}
		  \begin{axis} [
      axis lines=left,
      xlabel={$C_q$},
      ylabel={$P(q=1)$},
      ymin=0, ymax=1,
      xmin=-2, xmax=2,
      minor x tick num=3,
      minor y tick num=3,
      legend cell align=right,
      legend style={ draw=none, at={(1.3,1.0)}, },
    ]
    \addplot[red]    table[ x index=1, y index=2,  ]                {\plotdata};
    \addplot[gray]   table[ x index=1, y index=3,  forget plot, ]   {\plotdata};
    \addplot[cyan]   table[ x index=1, y index=4,  ]                {\plotdata};
    \addplot[gray]   table[ x index=1, y index=5,  forget plot, ]   {\plotdata};
    \addplot[violet] table[ x index=1, y index=6,  ]                {\plotdata};
    \addplot[gray]   table[ x index=1, y index=7,  forget plot, ]   {\plotdata};
    \addplot[teal]   table[ x index=1, y index=8,  ]                {\plotdata};
    \addplot[gray]   table[ x index=1, y index=9,  forget plot, ]   {\plotdata};
    \addplot[blue]   table[ x index=1, y index=10, ]                {\plotdata};
    \addplot[gray]   table[ x index=1, y index=11, forget plot, ]   {\plotdata};
    \addplot[orange] table[ x index=1, y index=12, ]                {\plotdata};
    \addplot[gray]   table[ x index=1, y index=13, forget plot, ]   {\plotdata};
    \addplot[yellow] table[ x index=1, y index=14, ]                {\plotdata};
    \addplot[gray]   table[ x index=1, y index=15, forget plot, ]   {\plotdata};
    \addplot[magenta]table[ x index=1, y index=16, ]                {\plotdata};
    \addplot[gray]   table[ x index=1, y index=17, forget plot, ]   {\plotdata};
    \addplot[green]  table[ x index=1, y index=18, ]                {\plotdata};
    \legend{-1.00,-0.75,-0.50,-0.25,0.00,0.25,0.50,0.75,1.00}
  \end{axis}
	    \end{tikzpicture}
	}
	\caption{ } \label{fig:vCHisingDcq}
    \end{subfigure}
    \hspace*{\fill} % separation between the subfigures
    \begin{subfigure}{0.49\textwidth}
	\scalebox{0.75}{
	    \begin{tikzpicture}
		\input{tab/chain_ISING_theoretic_vote_129_017}
		  \begin{axis} [
      axis lines=left,
      xlabel={$C_q$},
      ylabel={$P(q=1)$},
      ymin=0, ymax=1,
      xmin=-2, xmax=2,
      minor x tick num=3,
      minor y tick num=3,
      legend cell align=right,
      legend style={ draw=none, at={(1.3,1.0)}, },
    ]
    \addplot[red]    table[ x index=1, y index=2,  ]                {\plotdata};
    \addplot[gray]   table[ x index=1, y index=3,  forget plot, ]   {\plotdata};
    \addplot[cyan]   table[ x index=1, y index=4,  ]                {\plotdata};
    \addplot[gray]   table[ x index=1, y index=5,  forget plot, ]   {\plotdata};
    \addplot[violet] table[ x index=1, y index=6,  ]                {\plotdata};
    \addplot[gray]   table[ x index=1, y index=7,  forget plot, ]   {\plotdata};
    \addplot[teal]   table[ x index=1, y index=8,  ]                {\plotdata};
    \addplot[gray]   table[ x index=1, y index=9,  forget plot, ]   {\plotdata};
    \addplot[blue]   table[ x index=1, y index=10, ]                {\plotdata};
    \addplot[gray]   table[ x index=1, y index=11, forget plot, ]   {\plotdata};
    \addplot[orange] table[ x index=1, y index=12, ]                {\plotdata};
    \addplot[gray]   table[ x index=1, y index=13, forget plot, ]   {\plotdata};
    \addplot[yellow] table[ x index=1, y index=14, ]                {\plotdata};
    \addplot[gray]   table[ x index=1, y index=15, forget plot, ]   {\plotdata};
    \addplot[magenta]table[ x index=1, y index=16, ]                {\plotdata};
    \addplot[gray]   table[ x index=1, y index=17, forget plot, ]   {\plotdata};
    \addplot[green]  table[ x index=1, y index=18, ]                {\plotdata};
    \legend{-1.00,-0.75,-0.50,-0.25,0.00,0.25,0.50,0.75,1.00}
  \end{axis}
	    \end{tikzpicture}
	}
	\caption{ } \label{fig:vCHisingTcq}
    \end{subfigure}
    \caption{Plots of $P(q=1)$ vs. $C_q$ using the Ising model for 12 qubit chains (a) on AC12 and (b) theoretically. (17 different values of $C_c$ were plotted.)  }
    \label{fig:vCHisingCQ}
\end{figure}

\begin{figure}
    \begin{subfigure}{0.49\textwidth}
	\scalebox{0.75}{
	    \begin{tikzpicture}
		\input{tab/chain_QUBO_dwave_vote_129_017}
		  \begin{axis} [
      axis lines=left,
      xlabel={$C_q$},
      ylabel={$P(q=1)$},
      ymin=0, ymax=1,
      xmin=-2, xmax=2,
      minor x tick num=3,
      minor y tick num=3,
      legend cell align=right,
      legend style={ draw=none, at={(1.3,1.0)}, },
    ]
    \addplot[red]    table[ x index=1, y index=2,  ]                {\plotdata};
    \addplot[gray]   table[ x index=1, y index=3,  forget plot, ]   {\plotdata};
    \addplot[cyan]   table[ x index=1, y index=4,  ]                {\plotdata};
    \addplot[gray]   table[ x index=1, y index=5,  forget plot, ]   {\plotdata};
    \addplot[violet] table[ x index=1, y index=6,  ]                {\plotdata};
    \addplot[gray]   table[ x index=1, y index=7,  forget plot, ]   {\plotdata};
    \addplot[teal]   table[ x index=1, y index=8,  ]                {\plotdata};
    \addplot[gray]   table[ x index=1, y index=9,  forget plot, ]   {\plotdata};
    \addplot[blue]   table[ x index=1, y index=10, ]                {\plotdata};
    \addplot[gray]   table[ x index=1, y index=11, forget plot, ]   {\plotdata};
    \addplot[orange] table[ x index=1, y index=12, ]                {\plotdata};
    \addplot[gray]   table[ x index=1, y index=13, forget plot, ]   {\plotdata};
    \addplot[yellow] table[ x index=1, y index=14, ]                {\plotdata};
    \addplot[gray]   table[ x index=1, y index=15, forget plot, ]   {\plotdata};
    \addplot[magenta]table[ x index=1, y index=16, ]                {\plotdata};
    \addplot[gray]   table[ x index=1, y index=17, forget plot, ]   {\plotdata};
    \addplot[green]  table[ x index=1, y index=18, ]                {\plotdata};
    \legend{-1.00,-0.75,-0.50,-0.25,0.00,0.25,0.50,0.75,1.00}
  \end{axis}
	    \end{tikzpicture}
	}
	\caption{ } \label{fig:vCHquboDcq}
    \end{subfigure}
    \hspace*{\fill} % separation between the subfigures
    \begin{subfigure}{0.49\textwidth}
	\scalebox{0.75}{
	    \begin{tikzpicture}
		\input{tab/chain_QUBO_theoretic_vote_129_017}
		  \begin{axis} [
      axis lines=left,
      xlabel={$C_q$},
      ylabel={$P(q=1)$},
      ymin=0, ymax=1,
      xmin=-2, xmax=2,
      minor x tick num=3,
      minor y tick num=3,
      legend cell align=right,
      legend style={ draw=none, at={(1.3,1.0)}, },
    ]
    \addplot[red]    table[ x index=1, y index=2,  ]                {\plotdata};
    \addplot[gray]   table[ x index=1, y index=3,  forget plot, ]   {\plotdata};
    \addplot[cyan]   table[ x index=1, y index=4,  ]                {\plotdata};
    \addplot[gray]   table[ x index=1, y index=5,  forget plot, ]   {\plotdata};
    \addplot[violet] table[ x index=1, y index=6,  ]                {\plotdata};
    \addplot[gray]   table[ x index=1, y index=7,  forget plot, ]   {\plotdata};
    \addplot[teal]   table[ x index=1, y index=8,  ]                {\plotdata};
    \addplot[gray]   table[ x index=1, y index=9,  forget plot, ]   {\plotdata};
    \addplot[blue]   table[ x index=1, y index=10, ]                {\plotdata};
    \addplot[gray]   table[ x index=1, y index=11, forget plot, ]   {\plotdata};
    \addplot[orange] table[ x index=1, y index=12, ]                {\plotdata};
    \addplot[gray]   table[ x index=1, y index=13, forget plot, ]   {\plotdata};
    \addplot[yellow] table[ x index=1, y index=14, ]                {\plotdata};
    \addplot[gray]   table[ x index=1, y index=15, forget plot, ]   {\plotdata};
    \addplot[magenta]table[ x index=1, y index=16, ]                {\plotdata};
    \addplot[gray]   table[ x index=1, y index=17, forget plot, ]   {\plotdata};
    \addplot[green]  table[ x index=1, y index=18, ]                {\plotdata};
    \legend{-1.00,-0.75,-0.50,-0.25,0.00,0.25,0.50,0.75,1.00}
  \end{axis}
	    \end{tikzpicture}
	}
	\caption{ } \label{fig:vCHquboTcq}
    \end{subfigure}
    \caption{Plots of $P(q=1)$ vs. $C_q$ using the QUBO model for 12 qubit chains (a) on AC12 and (b) theoretically. (17 different values of $C_c$ were plotted.)  }
    \label{fig:vCHquboCQ}
\end{figure}

\begin{figure}
    \begin{subfigure}{0.49\textwidth}
	\scalebox{0.75}{
	    \begin{tikzpicture}
		\input{tab/chain_ISING_dwave_vote_017_129}
		  \begin{axis} [
      axis lines=left,
      xlabel={$C_c$},
      ylabel={$P(q=1)$},
      ymin=0, ymax=1,
      xmin=-1, xmax=1,
      minor x tick num=3,
      minor y tick num=3,
      legend cell align=right,
      legend style={ draw=none, at={(1.3,1.0)}, },
    ]
    \addplot[red]    table[ x index=1, y index=2,  ]                {\plotdata};
    \addplot[gray]   table[ x index=1, y index=3,  forget plot, ]   {\plotdata};
    \addplot[cyan]   table[ x index=1, y index=4,  ]                {\plotdata};
    \addplot[gray]   table[ x index=1, y index=5,  forget plot, ]   {\plotdata};
    \addplot[violet] table[ x index=1, y index=6,  ]                {\plotdata};
    \addplot[gray]   table[ x index=1, y index=7,  forget plot, ]   {\plotdata};
    \addplot[teal]   table[ x index=1, y index=8,  ]                {\plotdata};
    \addplot[gray]   table[ x index=1, y index=9,  forget plot, ]   {\plotdata};
    \addplot[blue]   table[ x index=1, y index=10, ]                {\plotdata};
    \addplot[gray]   table[ x index=1, y index=11, forget plot, ]   {\plotdata};
    \addplot[orange] table[ x index=1, y index=12, ]                {\plotdata};
    \addplot[gray]   table[ x index=1, y index=13, forget plot, ]   {\plotdata};
    \addplot[yellow] table[ x index=1, y index=14, ]                {\plotdata};
    \addplot[gray]   table[ x index=1, y index=15, forget plot, ]   {\plotdata};
    \addplot[magenta]table[ x index=1, y index=16, ]                {\plotdata};
    \addplot[gray]   table[ x index=1, y index=17, forget plot, ]   {\plotdata};
    \addplot[green]  table[ x index=1, y index=18, ]                {\plotdata};
    \legend{-2.00,-1.50,-1.00,-0.50,0.00,0.50,1.00,1.50,2.00}
  \end{axis}
	    \end{tikzpicture}
	}
	\caption{ } \label{fig:vCHisingDqc}
    \end{subfigure}
    \hspace*{\fill} % separation between the subfigures
    \begin{subfigure}{0.49\textwidth}
	\scalebox{0.75}{
	    \begin{tikzpicture}
		\input{tab/chain_ISING_theoretic_vote_017_129}
		  \begin{axis} [
      axis lines=left,
      xlabel={$C_c$},
      ylabel={$P(q=1)$},
      ymin=0, ymax=1,
      xmin=-1, xmax=1,
      minor x tick num=3,
      minor y tick num=3,
      legend cell align=right,
      legend style={ draw=none, at={(1.3,1.0)}, },
    ]
    \addplot[red]    table[ x index=1, y index=2,  ]                {\plotdata};
    \addplot[gray]   table[ x index=1, y index=3,  forget plot, ]   {\plotdata};
    \addplot[cyan]   table[ x index=1, y index=4,  ]                {\plotdata};
    \addplot[gray]   table[ x index=1, y index=5,  forget plot, ]   {\plotdata};
    \addplot[violet] table[ x index=1, y index=6,  ]                {\plotdata};
    \addplot[gray]   table[ x index=1, y index=7,  forget plot, ]   {\plotdata};
    \addplot[teal]   table[ x index=1, y index=8,  ]                {\plotdata};
    \addplot[gray]   table[ x index=1, y index=9,  forget plot, ]   {\plotdata};
    \addplot[blue]   table[ x index=1, y index=10, ]                {\plotdata};
    \addplot[gray]   table[ x index=1, y index=11, forget plot, ]   {\plotdata};
    \addplot[orange] table[ x index=1, y index=12, ]                {\plotdata};
    \addplot[gray]   table[ x index=1, y index=13, forget plot, ]   {\plotdata};
    \addplot[yellow] table[ x index=1, y index=14, ]                {\plotdata};
    \addplot[gray]   table[ x index=1, y index=15, forget plot, ]   {\plotdata};
    \addplot[magenta]table[ x index=1, y index=16, ]                {\plotdata};
    \addplot[gray]   table[ x index=1, y index=17, forget plot, ]   {\plotdata};
    \addplot[green]  table[ x index=1, y index=18, ]                {\plotdata};
    \legend{-2.00,-1.50,-1.00,-0.50,0.00,0.50,1.00,1.50,2.00}
  \end{axis}
	    \end{tikzpicture}
	}
	\caption{ } \label{fig:vCHisingTqc}
    \end{subfigure}
    \caption{Plots of $P(q=1)$ vs. $C_c$ using the Ising model for 12 qubit chains (a) on AC12 and (b) theoretically. (17 different values of $C_q$ were plotted.)  }
    \label{fig:vCHisingQC}
\end{figure}

\begin{figure}
    \begin{subfigure}{0.49\textwidth}
	\scalebox{0.75}{
	    \begin{tikzpicture}
		\input{tab/chain_QUBO_dwave_vote_017_129}
		  \begin{axis} [
      axis lines=left,
      xlabel={$C_c$},
      ylabel={$P(q=1)$},
      ymin=0, ymax=1,
      xmin=-1, xmax=1,
      minor x tick num=3,
      minor y tick num=3,
      legend cell align=right,
      legend style={ draw=none, at={(1.3,1.0)}, },
    ]
    \addplot[red]    table[ x index=1, y index=2,  ]                {\plotdata};
    \addplot[gray]   table[ x index=1, y index=3,  forget plot, ]   {\plotdata};
    \addplot[cyan]   table[ x index=1, y index=4,  ]                {\plotdata};
    \addplot[gray]   table[ x index=1, y index=5,  forget plot, ]   {\plotdata};
    \addplot[violet] table[ x index=1, y index=6,  ]                {\plotdata};
    \addplot[gray]   table[ x index=1, y index=7,  forget plot, ]   {\plotdata};
    \addplot[teal]   table[ x index=1, y index=8,  ]                {\plotdata};
    \addplot[gray]   table[ x index=1, y index=9,  forget plot, ]   {\plotdata};
    \addplot[blue]   table[ x index=1, y index=10, ]                {\plotdata};
    \addplot[gray]   table[ x index=1, y index=11, forget plot, ]   {\plotdata};
    \addplot[orange] table[ x index=1, y index=12, ]                {\plotdata};
    \addplot[gray]   table[ x index=1, y index=13, forget plot, ]   {\plotdata};
    \addplot[yellow] table[ x index=1, y index=14, ]                {\plotdata};
    \addplot[gray]   table[ x index=1, y index=15, forget plot, ]   {\plotdata};
    \addplot[magenta]table[ x index=1, y index=16, ]                {\plotdata};
    \addplot[gray]   table[ x index=1, y index=17, forget plot, ]   {\plotdata};
    \addplot[green]  table[ x index=1, y index=18, ]                {\plotdata};
    \legend{-2.00,-1.50,-1.00,-0.50,0.00,0.50,1.00,1.50,2.00}
  \end{axis}
	    \end{tikzpicture}
	}
	\caption{ } \label{fig:vCHquboDqc}
    \end{subfigure}
    \hspace*{\fill} % separation between the subfigures
    \begin{subfigure}{0.49\textwidth}
	\scalebox{0.75}{
	    \begin{tikzpicture}
		\input{tab/chain_QUBO_theoretic_vote_017_129}
		  \begin{axis} [
      axis lines=left,
      xlabel={$C_c$},
      ylabel={$P(q=1)$},
      ymin=0, ymax=1,
      xmin=-1, xmax=1,
      minor x tick num=3,
      minor y tick num=3,
      legend cell align=right,
      legend style={ draw=none, at={(1.3,1.0)}, },
    ]
    \addplot[red]    table[ x index=1, y index=2,  ]                {\plotdata};
    \addplot[gray]   table[ x index=1, y index=3,  forget plot, ]   {\plotdata};
    \addplot[cyan]   table[ x index=1, y index=4,  ]                {\plotdata};
    \addplot[gray]   table[ x index=1, y index=5,  forget plot, ]   {\plotdata};
    \addplot[violet] table[ x index=1, y index=6,  ]                {\plotdata};
    \addplot[gray]   table[ x index=1, y index=7,  forget plot, ]   {\plotdata};
    \addplot[teal]   table[ x index=1, y index=8,  ]                {\plotdata};
    \addplot[gray]   table[ x index=1, y index=9,  forget plot, ]   {\plotdata};
    \addplot[blue]   table[ x index=1, y index=10, ]                {\plotdata};
    \addplot[gray]   table[ x index=1, y index=11, forget plot, ]   {\plotdata};
    \addplot[orange] table[ x index=1, y index=12, ]                {\plotdata};
    \addplot[gray]   table[ x index=1, y index=13, forget plot, ]   {\plotdata};
    \addplot[yellow] table[ x index=1, y index=14, ]                {\plotdata};
    \addplot[gray]   table[ x index=1, y index=15, forget plot, ]   {\plotdata};
    \addplot[magenta]table[ x index=1, y index=16, ]                {\plotdata};
    \addplot[gray]   table[ x index=1, y index=17, forget plot, ]   {\plotdata};
    \addplot[green]  table[ x index=1, y index=18, ]                {\plotdata};
    \legend{-2.00,-1.50,-1.00,-0.50,0.00,0.50,1.00,1.50,2.00}
  \end{axis}
	    \end{tikzpicture}
	}
	\caption{ } \label{fig:vCHquboTqc}
    \end{subfigure}
    \caption{Plots of $P(q=1)$ vs. $C_c$ using the QUBO model for 12 qubit chains (a) on AC12 and (b) theoretically. (17 different values of $C_q$ were plotted.)  }
    \label{fig:vCHquboQC}
\end{figure}

\begin{figure}
    \begin{subfigure}{0.49\textwidth}
	\scalebox{0.75}{
	    \begin{tikzpicture}
		\input{tab/cell_ISING_dwave_vote_129_017}
		  \begin{axis} [
      axis lines=left,
      xlabel={$C_q$},
      ylabel={$P(q=1)$},
      ymin=0, ymax=1,
      xmin=-2, xmax=2,
      minor x tick num=3,
      minor y tick num=3,
      legend cell align=right,
      legend style={ draw=none, at={(1.3,1.0)}, },
    ]
    \addplot[red]    table[ x index=1, y index=2,  ]                {\plotdata};
    \addplot[gray]   table[ x index=1, y index=3,  forget plot, ]   {\plotdata};
    \addplot[cyan]   table[ x index=1, y index=4,  ]                {\plotdata};
    \addplot[gray]   table[ x index=1, y index=5,  forget plot, ]   {\plotdata};
    \addplot[violet] table[ x index=1, y index=6,  ]                {\plotdata};
    \addplot[gray]   table[ x index=1, y index=7,  forget plot, ]   {\plotdata};
    \addplot[teal]   table[ x index=1, y index=8,  ]                {\plotdata};
    \addplot[gray]   table[ x index=1, y index=9,  forget plot, ]   {\plotdata};
    \addplot[blue]   table[ x index=1, y index=10, ]                {\plotdata};
    \addplot[gray]   table[ x index=1, y index=11, forget plot, ]   {\plotdata};
    \addplot[orange] table[ x index=1, y index=12, ]                {\plotdata};
    \addplot[gray]   table[ x index=1, y index=13, forget plot, ]   {\plotdata};
    \addplot[yellow] table[ x index=1, y index=14, ]                {\plotdata};
    \addplot[gray]   table[ x index=1, y index=15, forget plot, ]   {\plotdata};
    \addplot[magenta]table[ x index=1, y index=16, ]                {\plotdata};
    \addplot[gray]   table[ x index=1, y index=17, forget plot, ]   {\plotdata};
    \addplot[green]  table[ x index=1, y index=18, ]                {\plotdata};
    \legend{-1.00,-0.75,-0.50,-0.25,0.00,0.25,0.50,0.75,1.00}
  \end{axis}
	    \end{tikzpicture}
	}
	\caption{ } \label{fig:vCEisingDcq}
    \end{subfigure}
    \hspace*{\fill} % separation between the subfigures
    \begin{subfigure}{0.49\textwidth}
	\scalebox{0.75}{
	    \begin{tikzpicture}
		\input{tab/cell_ISING_theoretic_vote_129_017}
		  \begin{axis} [
      axis lines=left,
      xlabel={$C_q$},
      ylabel={$P(q=1)$},
      ymin=0, ymax=1,
      xmin=-2, xmax=2,
      minor x tick num=3,
      minor y tick num=3,
      legend cell align=right,
      legend style={ draw=none, at={(1.3,1.0)}, },
    ]
    \addplot[red]    table[ x index=1, y index=2,  ]                {\plotdata};
    \addplot[gray]   table[ x index=1, y index=3,  forget plot, ]   {\plotdata};
    \addplot[cyan]   table[ x index=1, y index=4,  ]                {\plotdata};
    \addplot[gray]   table[ x index=1, y index=5,  forget plot, ]   {\plotdata};
    \addplot[violet] table[ x index=1, y index=6,  ]                {\plotdata};
    \addplot[gray]   table[ x index=1, y index=7,  forget plot, ]   {\plotdata};
    \addplot[teal]   table[ x index=1, y index=8,  ]                {\plotdata};
    \addplot[gray]   table[ x index=1, y index=9,  forget plot, ]   {\plotdata};
    \addplot[blue]   table[ x index=1, y index=10, ]                {\plotdata};
    \addplot[gray]   table[ x index=1, y index=11, forget plot, ]   {\plotdata};
    \addplot[orange] table[ x index=1, y index=12, ]                {\plotdata};
    \addplot[gray]   table[ x index=1, y index=13, forget plot, ]   {\plotdata};
    \addplot[yellow] table[ x index=1, y index=14, ]                {\plotdata};
    \addplot[gray]   table[ x index=1, y index=15, forget plot, ]   {\plotdata};
    \addplot[magenta]table[ x index=1, y index=16, ]                {\plotdata};
    \addplot[gray]   table[ x index=1, y index=17, forget plot, ]   {\plotdata};
    \addplot[green]  table[ x index=1, y index=18, ]                {\plotdata};
    \legend{-1.00,-0.75,-0.50,-0.25,0.00,0.25,0.50,0.75,1.00}
  \end{axis}
	    \end{tikzpicture}
	}
	\caption{ } \label{fig:vCEisingTcq}
    \end{subfigure}
    \caption{Plots of $P(q=1)$ vs. $C_q$ using the Ising model for 8 qubit cells (a) on AC12 and (b) theoretically. (17 different values of $C_c$ were plotted.)  }
    \label{fig:vCEisingCQ}
\end{figure}

\begin{figure}
    \begin{subfigure}{0.49\textwidth}
	\scalebox{0.75}{
	    \begin{tikzpicture}
		\input{tab/cell_QUBO_dwave_vote_129_017}
		  \begin{axis} [
      axis lines=left,
      xlabel={$C_q$},
      ylabel={$P(q=1)$},
      ymin=0, ymax=1,
      xmin=-2, xmax=2,
      minor x tick num=3,
      minor y tick num=3,
      legend cell align=right,
      legend style={ draw=none, at={(1.3,1.0)}, },
    ]
    \addplot[red]    table[ x index=1, y index=2,  ]                {\plotdata};
    \addplot[gray]   table[ x index=1, y index=3,  forget plot, ]   {\plotdata};
    \addplot[cyan]   table[ x index=1, y index=4,  ]                {\plotdata};
    \addplot[gray]   table[ x index=1, y index=5,  forget plot, ]   {\plotdata};
    \addplot[violet] table[ x index=1, y index=6,  ]                {\plotdata};
    \addplot[gray]   table[ x index=1, y index=7,  forget plot, ]   {\plotdata};
    \addplot[teal]   table[ x index=1, y index=8,  ]                {\plotdata};
    \addplot[gray]   table[ x index=1, y index=9,  forget plot, ]   {\plotdata};
    \addplot[blue]   table[ x index=1, y index=10, ]                {\plotdata};
    \addplot[gray]   table[ x index=1, y index=11, forget plot, ]   {\plotdata};
    \addplot[orange] table[ x index=1, y index=12, ]                {\plotdata};
    \addplot[gray]   table[ x index=1, y index=13, forget plot, ]   {\plotdata};
    \addplot[yellow] table[ x index=1, y index=14, ]                {\plotdata};
    \addplot[gray]   table[ x index=1, y index=15, forget plot, ]   {\plotdata};
    \addplot[magenta]table[ x index=1, y index=16, ]                {\plotdata};
    \addplot[gray]   table[ x index=1, y index=17, forget plot, ]   {\plotdata};
    \addplot[green]  table[ x index=1, y index=18, ]                {\plotdata};
    \legend{-1.00,-0.75,-0.50,-0.25,0.00,0.25,0.50,0.75,1.00}
  \end{axis}
	    \end{tikzpicture}
	}
	\caption{ } \label{fig:vCEquboDcq}
    \end{subfigure}
    \hspace*{\fill} % separation between the subfigures
    \begin{subfigure}{0.49\textwidth}
	\scalebox{0.75}{
	    \begin{tikzpicture}
		\input{tab/cell_QUBO_theoretic_vote_129_017}
		  \begin{axis} [
      axis lines=left,
      xlabel={$C_q$},
      ylabel={$P(q=1)$},
      ymin=0, ymax=1,
      xmin=-2, xmax=2,
      minor x tick num=3,
      minor y tick num=3,
      legend cell align=right,
      legend style={ draw=none, at={(1.3,1.0)}, },
    ]
    \addplot[red]    table[ x index=1, y index=2,  ]                {\plotdata};
    \addplot[gray]   table[ x index=1, y index=3,  forget plot, ]   {\plotdata};
    \addplot[cyan]   table[ x index=1, y index=4,  ]                {\plotdata};
    \addplot[gray]   table[ x index=1, y index=5,  forget plot, ]   {\plotdata};
    \addplot[violet] table[ x index=1, y index=6,  ]                {\plotdata};
    \addplot[gray]   table[ x index=1, y index=7,  forget plot, ]   {\plotdata};
    \addplot[teal]   table[ x index=1, y index=8,  ]                {\plotdata};
    \addplot[gray]   table[ x index=1, y index=9,  forget plot, ]   {\plotdata};
    \addplot[blue]   table[ x index=1, y index=10, ]                {\plotdata};
    \addplot[gray]   table[ x index=1, y index=11, forget plot, ]   {\plotdata};
    \addplot[orange] table[ x index=1, y index=12, ]                {\plotdata};
    \addplot[gray]   table[ x index=1, y index=13, forget plot, ]   {\plotdata};
    \addplot[yellow] table[ x index=1, y index=14, ]                {\plotdata};
    \addplot[gray]   table[ x index=1, y index=15, forget plot, ]   {\plotdata};
    \addplot[magenta]table[ x index=1, y index=16, ]                {\plotdata};
    \addplot[gray]   table[ x index=1, y index=17, forget plot, ]   {\plotdata};
    \addplot[green]  table[ x index=1, y index=18, ]                {\plotdata};
    \legend{-1.00,-0.75,-0.50,-0.25,0.00,0.25,0.50,0.75,1.00}
  \end{axis}
	    \end{tikzpicture}
	}
	\caption{ } \label{fig:vCEquboTcq}
    \end{subfigure}
    \caption{Plots of $P(q=1)$ vs. $C_q$ using the QUBO model for 8 qubit cells (a) on AC12 and (b) theoretically. (17 different values of $C_c$ were plotted.)  }
    \label{fig:vCEquboCQ}
\end{figure}

\begin{figure}
    \begin{subfigure}{0.49\textwidth}
	\scalebox{0.75}{
	    \begin{tikzpicture}
		\input{tab/cell_ISING_dwave_vote_017_129}
		  \begin{axis} [
      axis lines=left,
      xlabel={$C_c$},
      ylabel={$P(q=1)$},
      ymin=0, ymax=1,
      xmin=-1, xmax=1,
      minor x tick num=3,
      minor y tick num=3,
      legend cell align=right,
      legend style={ draw=none, at={(1.3,1.0)}, },
    ]
    \addplot[red]    table[ x index=1, y index=2,  ]                {\plotdata};
    \addplot[gray]   table[ x index=1, y index=3,  forget plot, ]   {\plotdata};
    \addplot[cyan]   table[ x index=1, y index=4,  ]                {\plotdata};
    \addplot[gray]   table[ x index=1, y index=5,  forget plot, ]   {\plotdata};
    \addplot[violet] table[ x index=1, y index=6,  ]                {\plotdata};
    \addplot[gray]   table[ x index=1, y index=7,  forget plot, ]   {\plotdata};
    \addplot[teal]   table[ x index=1, y index=8,  ]                {\plotdata};
    \addplot[gray]   table[ x index=1, y index=9,  forget plot, ]   {\plotdata};
    \addplot[blue]   table[ x index=1, y index=10, ]                {\plotdata};
    \addplot[gray]   table[ x index=1, y index=11, forget plot, ]   {\plotdata};
    \addplot[orange] table[ x index=1, y index=12, ]                {\plotdata};
    \addplot[gray]   table[ x index=1, y index=13, forget plot, ]   {\plotdata};
    \addplot[yellow] table[ x index=1, y index=14, ]                {\plotdata};
    \addplot[gray]   table[ x index=1, y index=15, forget plot, ]   {\plotdata};
    \addplot[magenta]table[ x index=1, y index=16, ]                {\plotdata};
    \addplot[gray]   table[ x index=1, y index=17, forget plot, ]   {\plotdata};
    \addplot[green]  table[ x index=1, y index=18, ]                {\plotdata};
    \legend{-2.00,-1.50,-1.00,-0.50,0.00,0.50,1.00,1.50,2.00}
  \end{axis}
	    \end{tikzpicture}
	}
	\caption{ } \label{fig:vCEisingDqc}
    \end{subfigure}
    \hspace*{\fill} % separation between the subfigures
    \begin{subfigure}{0.49\textwidth}
	\scalebox{0.75}{
	    \begin{tikzpicture}
		\input{tab/cell_ISING_theoretic_vote_017_129}
		  \begin{axis} [
      axis lines=left,
      xlabel={$C_c$},
      ylabel={$P(q=1)$},
      ymin=0, ymax=1,
      xmin=-1, xmax=1,
      minor x tick num=3,
      minor y tick num=3,
      legend cell align=right,
      legend style={ draw=none, at={(1.3,1.0)}, },
    ]
    \addplot[red]    table[ x index=1, y index=2,  ]                {\plotdata};
    \addplot[gray]   table[ x index=1, y index=3,  forget plot, ]   {\plotdata};
    \addplot[cyan]   table[ x index=1, y index=4,  ]                {\plotdata};
    \addplot[gray]   table[ x index=1, y index=5,  forget plot, ]   {\plotdata};
    \addplot[violet] table[ x index=1, y index=6,  ]                {\plotdata};
    \addplot[gray]   table[ x index=1, y index=7,  forget plot, ]   {\plotdata};
    \addplot[teal]   table[ x index=1, y index=8,  ]                {\plotdata};
    \addplot[gray]   table[ x index=1, y index=9,  forget plot, ]   {\plotdata};
    \addplot[blue]   table[ x index=1, y index=10, ]                {\plotdata};
    \addplot[gray]   table[ x index=1, y index=11, forget plot, ]   {\plotdata};
    \addplot[orange] table[ x index=1, y index=12, ]                {\plotdata};
    \addplot[gray]   table[ x index=1, y index=13, forget plot, ]   {\plotdata};
    \addplot[yellow] table[ x index=1, y index=14, ]                {\plotdata};
    \addplot[gray]   table[ x index=1, y index=15, forget plot, ]   {\plotdata};
    \addplot[magenta]table[ x index=1, y index=16, ]                {\plotdata};
    \addplot[gray]   table[ x index=1, y index=17, forget plot, ]   {\plotdata};
    \addplot[green]  table[ x index=1, y index=18, ]                {\plotdata};
    \legend{-2.00,-1.50,-1.00,-0.50,0.00,0.50,1.00,1.50,2.00}
  \end{axis}
	    \end{tikzpicture}
	}
	\caption{ } \label{fig:vCEisingTqc}
    \end{subfigure}
    \caption{Plots of $P(q=1)$ vs. $C_c$ using the Ising model for 8 qubit cells (a) on AC12 and (b) theoretically. (17 different values of $C_q$ were plotted.)  }
    \label{fig:vCEisingQC}
\end{figure}

\afterpage{
\vspace*{-1cm}
\begin{figure}
    \begin{subfigure}{0.49\textwidth}
	\scalebox{0.75}{
	    \begin{tikzpicture}
		\input{tab/cell_QUBO_dwave_vote_017_129}
		  \begin{axis} [
      axis lines=left,
      xlabel={$C_c$},
      ylabel={$P(q=1)$},
      ymin=0, ymax=1,
      xmin=-1, xmax=1,
      minor x tick num=3,
      minor y tick num=3,
      legend cell align=right,
      legend style={ draw=none, at={(1.3,1.0)}, },
    ]
    \addplot[red]    table[ x index=1, y index=2,  ]                {\plotdata};
    \addplot[gray]   table[ x index=1, y index=3,  forget plot, ]   {\plotdata};
    \addplot[cyan]   table[ x index=1, y index=4,  ]                {\plotdata};
    \addplot[gray]   table[ x index=1, y index=5,  forget plot, ]   {\plotdata};
    \addplot[violet] table[ x index=1, y index=6,  ]                {\plotdata};
    \addplot[gray]   table[ x index=1, y index=7,  forget plot, ]   {\plotdata};
    \addplot[teal]   table[ x index=1, y index=8,  ]                {\plotdata};
    \addplot[gray]   table[ x index=1, y index=9,  forget plot, ]   {\plotdata};
    \addplot[blue]   table[ x index=1, y index=10, ]                {\plotdata};
    \addplot[gray]   table[ x index=1, y index=11, forget plot, ]   {\plotdata};
    \addplot[orange] table[ x index=1, y index=12, ]                {\plotdata};
    \addplot[gray]   table[ x index=1, y index=13, forget plot, ]   {\plotdata};
    \addplot[yellow] table[ x index=1, y index=14, ]                {\plotdata};
    \addplot[gray]   table[ x index=1, y index=15, forget plot, ]   {\plotdata};
    \addplot[magenta]table[ x index=1, y index=16, ]                {\plotdata};
    \addplot[gray]   table[ x index=1, y index=17, forget plot, ]   {\plotdata};
    \addplot[green]  table[ x index=1, y index=18, ]                {\plotdata};
    \legend{-2.00,-1.50,-1.00,-0.50,0.00,0.50,1.00,1.50,2.00}
  \end{axis}
	    \end{tikzpicture}
	}
	\caption{ } \label{fig:vCEquboDqc}
    \end{subfigure}
    \hspace*{\fill} % separation between the subfigures
    \begin{subfigure}{0.49\textwidth}
	\scalebox{0.75}{
	    \begin{tikzpicture}
		\input{tab/cell_QUBO_theoretic_vote_017_129}
		  \begin{axis} [
      axis lines=left,
      xlabel={$C_c$},
      ylabel={$P(q=1)$},
      ymin=0, ymax=1,
      xmin=-1, xmax=1,
      minor x tick num=3,
      minor y tick num=3,
      legend cell align=right,
      legend style={ draw=none, at={(1.3,1.0)}, },
    ]
    \addplot[red]    table[ x index=1, y index=2,  ]                {\plotdata};
    \addplot[gray]   table[ x index=1, y index=3,  forget plot, ]   {\plotdata};
    \addplot[cyan]   table[ x index=1, y index=4,  ]                {\plotdata};
    \addplot[gray]   table[ x index=1, y index=5,  forget plot, ]   {\plotdata};
    \addplot[violet] table[ x index=1, y index=6,  ]                {\plotdata};
    \addplot[gray]   table[ x index=1, y index=7,  forget plot, ]   {\plotdata};
    \addplot[teal]   table[ x index=1, y index=8,  ]                {\plotdata};
    \addplot[gray]   table[ x index=1, y index=9,  forget plot, ]   {\plotdata};
    \addplot[blue]   table[ x index=1, y index=10, ]                {\plotdata};
    \addplot[gray]   table[ x index=1, y index=11, forget plot, ]   {\plotdata};
    \addplot[orange] table[ x index=1, y index=12, ]                {\plotdata};
    \addplot[gray]   table[ x index=1, y index=13, forget plot, ]   {\plotdata};
    \addplot[yellow] table[ x index=1, y index=14, ]                {\plotdata};
    \addplot[gray]   table[ x index=1, y index=15, forget plot, ]   {\plotdata};
    \addplot[magenta]table[ x index=1, y index=16, ]                {\plotdata};
    \addplot[gray]   table[ x index=1, y index=17, forget plot, ]   {\plotdata};
    \addplot[green]  table[ x index=1, y index=18, ]                {\plotdata};
    \legend{-2.00,-1.50,-1.00,-0.50,0.00,0.50,1.00,1.50,2.00}
  \end{axis}
	    \end{tikzpicture}
	}
	\caption{ } \label{fig:vCEquboTqc}
    \end{subfigure}
    \caption{Plots of $P(q=1)$ vs. $C_c$ using the QUBO model for 8 qubit cells (a) on AC12 and (b) theoretically. (17 different values of $C_q$ were plotted.)  }
    \label{fig:vCEquboQC}
\end{figure}
} % end of "afterpage" group

\end{document}